\documentclass{aa}  

\usepackage{graphicx}
\usepackage{txfonts}

\begin{document}

   \title{Spectral and atmospheric characterisation of a new benchmark brown dwarf HD~13724~B \thanks{Based on observations collected with SPHERE mounted on the VLT at Paranal Observatory (ESO, Chile) under programmes 0102.C-0236(A) (PI: Rickman) and 0104.C-0702(B) (PI: Rickman) as well as observations collected with the CORALIE spectrograph mounted on the 1.2 m Swiss telescope at La Silla Observatory and with the HARPS spectrograph on the ESO 3.6 m telescope at La Silla (ESO, Chile).}
\thanks{The radial-velocity measurements, reduced images and additional data products discussed in this paper are available on the DACE web platform at https://dace.unige.ch/.}}

   \author{E. L. Rickman \inst{1}, D. S\'{e}gransan \inst{1}, J. Hagelberg \inst{1}, J.-L. Beuzit \inst{2}, A. Cheetham \inst{1}, J.-B. Delisle \inst{1}, T. Forveille \inst{3}, S. Udry \inst{1}}

   \institute{D\'{e}partment d'astronomie de l'Universit\'{e} de Gen\`{e}ve, 51 ch. des Maillettes Sauverny, 1290 Versoix, Switzerland\\
              \email{emily.rickman@unige.ch}
    \and
    Aix Marseille Univ., CNRS, CNES, LAM, Marseille, France
    \and
    Univ. Grenoble Alpes, CNRS, IPAG, 38000 Grenoble, France}
            
    \date{Received ; accepted }
    \authorrunning{Rickman et al.}
 
  \abstract
   {HD~13724 is a nearby solar-type star at 43.48 $\pm$ 0.06~pc hosting a long-period low-mass brown dwarf detected with the CORALIE echelle spectrograph as part of the historical CORALIE radial-velocity search for extra-solar planets. The companion has a minimum mass of $26.77^{+4.4}_{-2.2}~M_{\mathrm{Jup}}$ and an expected semi-major axis of $\sim$240 mas making it a suitable target for further characterisation with high-contrast imaging, in particular to measure its inclination, mass, and spectrum and thus establish its substellar nature.}
   {Using high-contrast imaging with the SPHERE instrument on the Very Large Telescope (VLT), we are able to directly image a brown dwarf companion to HD~13724 and obtain a low-resolution spectrum.}
   {We combine the radial-velocity measurements of CORALIE and HARPS taken over two decades and high-contrast imaging from SPHERE to obtain a dynamical mass estimate. From the SPHERE data we obtain a low-resolution spectrum of the companion from $Y$ to $J$ band, as well as photometric measurements from IRDIS in the $J$, $H,$ and $K$ bands.}
   {Using high-contrast imaging with the SPHERE instrument at the VLT, we report the first images of a brown dwarf companion orbiting the host star HD~13724. It has an angular separation of 175.6 $\pm$ 4.5 mas and an $H$-band contrast of $10.61\pm0.16$ mag, and using the age estimate of the star to be $\sim$1~Gyr gives an isochronal mass estimate of $\sim$44~$M_{\mathrm{Jup}}$. By combining radial-velocity and imaging data we also obtain a dynamical mass of $50.5^{+3.3}_{-3.5}~M_{\mathrm{Jup}}$. Through fitting an atmospheric model, we estimate a surface gravity of $\log g = 5.5$ and an effective temperature of 1000~K. A comparison of its spectrum with observed T dwarfs estimates a spectral type of T4 or T4.5, with a T4 object providing the best fit.}
   {}


   \maketitle
%

\section{Introduction}

Evolutionary models of brown dwarfs are plagued by a lack of observational constraints in addition to model degeneracies. The complex molecular chemistry of their atmospheres leaves a relatively wide parameter space for models to span. For this reason, the detection of brown dwarfs is vital to testing the complex atmospheres,  structure, and evolution of these substellar objects \citep{2003A&A...402..701B, 2015A&A...577A..42B}. Furthermore, it is important to characterise brown dwarfs that are orbiting stars, where the age of the system can be constrained, unlike field brown dwarfs. In addition, brown dwarfs orbiting stars provide an opportunity to monitor the radial velocity (RV) of these systems which allows us to place constraints on their dynamical masses over time \citep{2006ApJ...644.1193B}.

With over 20 years worth of RV measurements from the CORALIE survey for extrasolar planets \citep{2000fepc.conf..571U}, HD~13724 has been identified as a promising candidate for observational follow-up with direct imaging \citep{2019A&A...625A..71R} due to its minimum mass ($26.77^{+4.4}_{-2.2} M_{\mathrm{Jup}}$) and expected semi-major axis ($\sim$240~mas), which was calculated from the orbital period, itself derived from RV measurements.

The CORALIE RV survey is a volume-limited sample of 1647 main sequence stars from F8 down to K0 located within 50~pc of the Sun. Such a long base line of observations allows us to detect massive giant planets at separations of larger than 5~AU. This in turn identifies golden targets for direct imaging, as such companions are rare and are very difficult to search for blindly. Selecting long-period candidates to image from RV surveys has proven to be valuable (see \cite{2018A&A...614A..16C, 2019A&A...631A.107P}).

Radial-velocity measurements provide a lower limit on the measured masses due to the unknown orbital inclination. Therefore, directly imaging long-period RV candidates allows us to break that degeneracy and provide constraints on the dynamical mass of the companion.

Not only does combining these two detection techniques allow us to start filling in a largely unexplored parameters space, but through combining RV and direct imaging data we can now expect to dynamically measure the mass of such companions. By constraining the mass, we are able to place additional constraints on the evolution of the companion, both in terms of temperature and atmospheric composition (e.g. see \citet{2016A&A...587A..56M, 2016A&A...587A..55V}).

To date, individual dynamical masses from combining radial velocity and imaging measurements are known for only a handful of brown dwarfs \citep{2009ApJ...707L.123T,2011A&A...525A..95S,2012ApJ...751...97C, 2014ApJ...781...29C, 2016ApJ...831..136C, 2017ApJS..231...15D,2019A&A...631A.107P,2018AJ....155..159B,2018A&A...614A..16C,2019arXiv191001652B,2019arXiv191202565M}, therefore any new detections contribute significantly to brown dwarf models in addition to providing important analogues for the characterisation of exoplanets. This forms part of a larger effort to determine the giant planet upper mass limit and lower mass limit for brown dwarfs, especially in the $\sim$20~$M_{\mathrm{Jup}}-$40~$M_{\mathrm{Jup}}$ range where there is a dearth of  observed companions \citep{2011A&A...525A..95S}. Detecting brown dwarfs in this parameter space can help us to understand the formation of these objects, whether they formed via gravitational instability like binary systems or via core-accretion, as in the case of planets. This is crucial to understanding the formation processes of such systems and to defining the boundary between massive planets and low-mass brown dwarfs.

To determine the mass of an imaged brown dwarf companion, the key parameter for the evolution of substellar objects, we usually rely on evolutionary models \citep{1996Sci...272.1919M, 2003A&A...402..701B, 2012RSPTA.370.2765A, 2012ApJ...756..172M}. These models still need to be tested and properly calibrated through observations and the discovery of benchmark sources provides a powerful and critical tool to achieve this. Furthermore, as we move toward imaging increasingly small objects it is important to use them to test theoretical atmospheric models.

Typically, the conditions around  young stars are more favourable to direct imaging  because any companion will still be bright and hot and therefore easier to detect. In contrast, the RV method is typically suited to older stars where the RV signal is not too contaminated from variability caused by stellar activity in young and active stars. Consequently, combining these two techniques allows us to not only probe a mass-separation parameter space that is largely unexplored, but also bridge the gap between younger and older companion candidates.

Here, we report the first images and low-resolution spectrum of the benchmark brown dwarf HD~13724~B. In addition, we extend the time baseline of the RV observations. When combined with the imaging data, this allows constraints to be placed on the mass and orbital parameters of the brown dwarf companion. Thanks to the high precision of the RV data and the number of points, the minimum mass is well constrained, meaning that only a few astrometric points from high-contrast imaging were necessary to ensure a high-precision orbit and dynamical mass.

The paper is organised as follows. The properties of the host star  are summarised in Sect. 2. In Sect. 3 we summarise the observations and data analysis procedures for the RV and high-contrast imaging data. In Sect. 4 we present the results of the imaging data analysis and derived companion properties, and give an overview of the results from the combined orbital fitting. The results are discussed in Sect. 5 with some concluding remarks.

\section{Stellar characterisation}

The observed and inferred stellar parameters for HD~13724 are summarised in Table~\ref{table:1}. The spectral type, V band magnitude, and colour index are taken from the HIPPARCOS and Tycho catalogues \citep{1997A&A...323L..57H, 1997A&A...323L..49P}, while the astrometric parallax ($\pi$) and luminosity are taken from the second \emph{Gaia} data release \citep{2018A&A...616A...1G}. The effective temperature, gravity and metallicities were derived using the same spectroscopic methods as applied in \cite{2013A&A...556A.150S}, whilst the $v \sin i$ is computed using the calibration of CORALIE's Cross Correlation Function \citep[CCF;][]{2001A&A...373.1019S, marmier_phd_thesis}.

The mean chromospheric activity index - $\log(R'_{HK})$ - is computed by co-adding the corresponding CORALIE spectra to improve the signal-to-noise ratio which allows us to measure the Ca II re-emission at $\lambda = 3933.66$~\AA. We derived an estimate of the rotational period of the star from the mean $\log (R'_{HK})$ activity index using the calibration of \cite{2008ApJ...687.1264M}. 

The stellar radius and uncertainties are derived from the \emph{Gaia} luminosities and the effective temperatures are obtained from the spectroscopic analysis. A systematic error of 50~K was quadratically added to the effective temperature error bars and propagated in the radius uncertainties. 

The mass and age of HD~13724, as well as the uncertainties, were derived using the Geneva stellar evolution modes \citep{2012A&A...537A.146E, 2013A&A...558A.103G}. The interpolation in the model grid was made through a Bayesian formalism using observational Gaussian priors on $T_{\mathrm{eff}}$, $M_V$, $\log g,$ and [Fe/H] \citep{marmier_phd_thesis}.

\begin{table}
\caption{Stellar parameters of HD~13724}     
\label{table:1}      
\centering           
\begin{tabular}{c c c c}        
\hline\hline                    
Parameters & units & HD~13724 \\    
\hline                  
    Spectral type \tablefoottext{a} & & G3/G5V \\ 
    $V$ \tablefoottext{a} & & 7.89 \\
    $B-V$ \tablefoottext{a} & & 0.667 \\
    $\pi$ \tablefoottext{b} & $[\text{mas}]$ & 23.0 $\pm$ 0.03 \\
    $M_{v}$ & & 4.70 \\
    $T_{\mathrm{Gaia}}$ \tablefoottext{b} & $[\text{K}]$ & $5775.33^{+49.67}_{-74.33}$ \\
    $\log g$& $[\text{cgs}]$ & 4.44 $\pm$ 0.07 \\
    $[\text{Fe/H}]$ & $[\text{dex}]$ & 0.23 $\pm$ 0.02 \\
    $v \sin i$ \tablefoottext{c} & $[\text{km s}^{-1}]$ & 3.025 \\
    $M_{*}$ & $[\text{M}_{\odot}]$ & 1.14 $\pm$ 0.06 \\
    $L_{\mathrm{Gaia}}$ \tablefoottext{b} & $[L_{\odot}]$ & $1.14^{+0.001}_{-0.002}$ \\
    $R_{\mathrm{Gaia}}$ \tablefoottext{b} & $[R_{\odot}]$ & 1.07 $\pm$ 0.02 \\
    $\log R^{'}_{HK}$ \tablefoottext{c} & & -4.76 $\pm$ 0.003 \\
    $P_{\text{rot}}$ & $[\text{days}]$ & 20.2$\pm$1.2 \\
    Age & $[\text{Gyr}]$ & 1.04 $\pm$ 0.88 \\
\hline          
\end{tabular}
\tablefoot{
\tablefoottext{a}{Parameters taken from HIPPARCOS and Tycho \citep{1997A&A...323L..49P, 1997A&A...323L..57H}}
\tablefoottext{b}{Parallaxes taken from \emph{Gaia} data release 2 \citep{2018A&A...616A...1G}}
\tablefoottext{c}{Parameters derived using CORALIE CCF.}}
\end{table}

\section{Observations and data reduction}

Radial-velocity and direct imaging observations were combined to constrain the orbit of HD~13724~B. Furthermore, the extensive orbital coverage of the RV time series allows us to precisely constrain its orbital parameters. Combined with several direct imaging observations we are able to derive the orbital inclination $i$ and thus the mass. In addition, the SPHERE high-contrast IRDIS observations provide six narrow-band-width photometric measurements in the $J$, $H,$ and $K$ bands and IFS observations allow us to obtain a low-resolution spectrum of the brown dwarf companion in the $Y-J$ bands.

\subsection{Radial velocities}{\label{sec:3.1}}

HD~13724 has been observed since August 1999 with the CORALIE spectrograph \citep{2000fepc.conf..548Q} installed on the 1.2m EULER Swiss telescope at La Silla observatory (Chile) and HARPS \citep{2003Msngr.114...20M} on the ESO/3.6~m telescope to obtain RVs. 

\begin{figure}
 \includegraphics[width=0.5\textwidth]{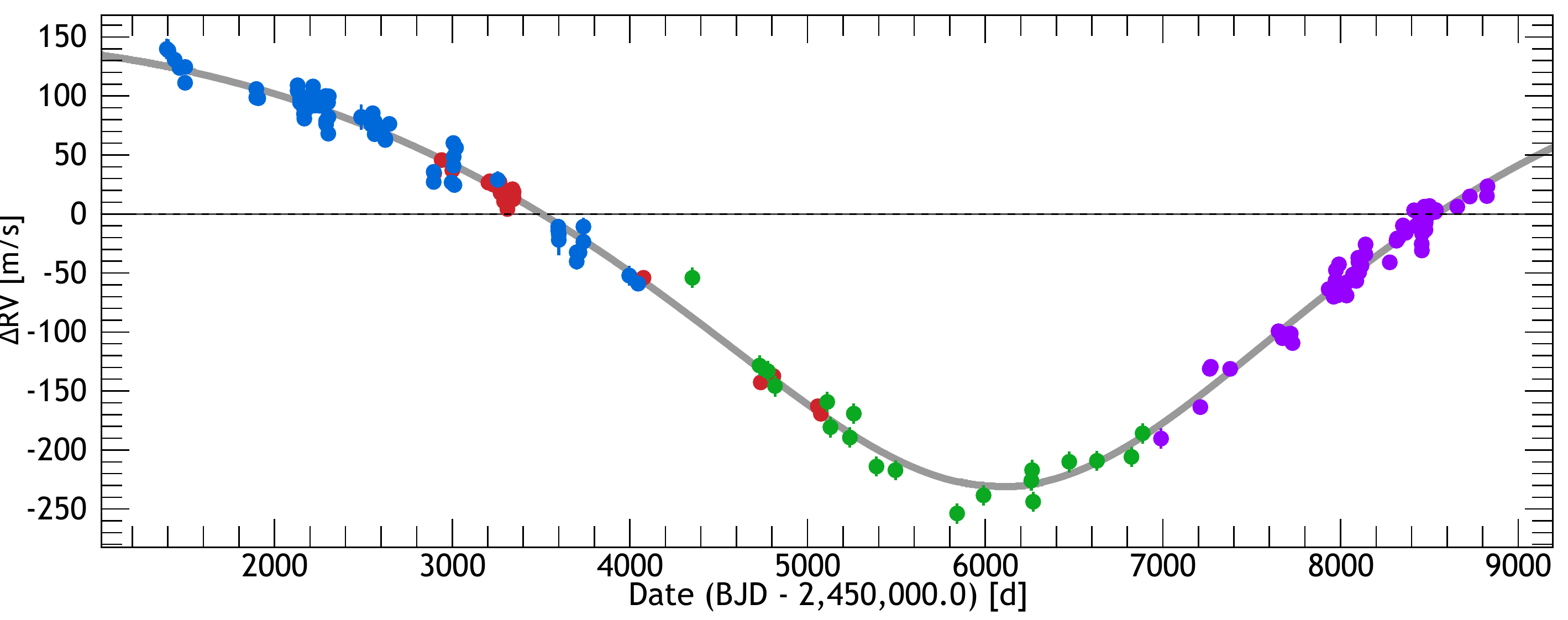}
  \caption{HD13724 RV measurements as a function of Julian Dates obtained with CORALIE-98 (blue), CORALIE-07 (green), and CORALIE-14 (purple). The HARPS-03 data are shown in red. The best single-planet Keplerian model is represented as a black curve.}
  \label{fig:1}
\end{figure}

The RV data analysis presented in this paper was accomplished using a set of online tools hosted by the Data and Analysis Center for Exoplanets (DACE) \footnote{The DACE platform is available at \url{https://dace.unige.ch} where the online tools to analyse RV data can be found in the section Observations->Radial Velocities.}, which performs a Keplerian fit to the data as described in \cite{2016A&A...590A.134D}. The 179 measurements obtained between August 1999 and December 2019 are shown in Fig.~\ref{fig:1} with the corresponding Keplerian model.

Following on from \cite{2019A&A...625A..71R} we obtained five more RV measurements over $\sim$10 months. We also added four historical and low-precision ($\sim$ 300~m/s) CORAVEL measurements to increase the overall time-span of the RV measurements by 10 years and to provide an upper bound constraint on the minimum mass of the companion. The RV data products presented in this paper are available at DACE with the new orbital parameters shown in Table~\ref{tab:ds1}. The full description of the RV data is outlined in \cite{2019A&A...625A..71R}.

\subsection{SPHERE high-contrast imaging}

\begin{figure*}
\begin{center}
\includegraphics[width=.28\textwidth]{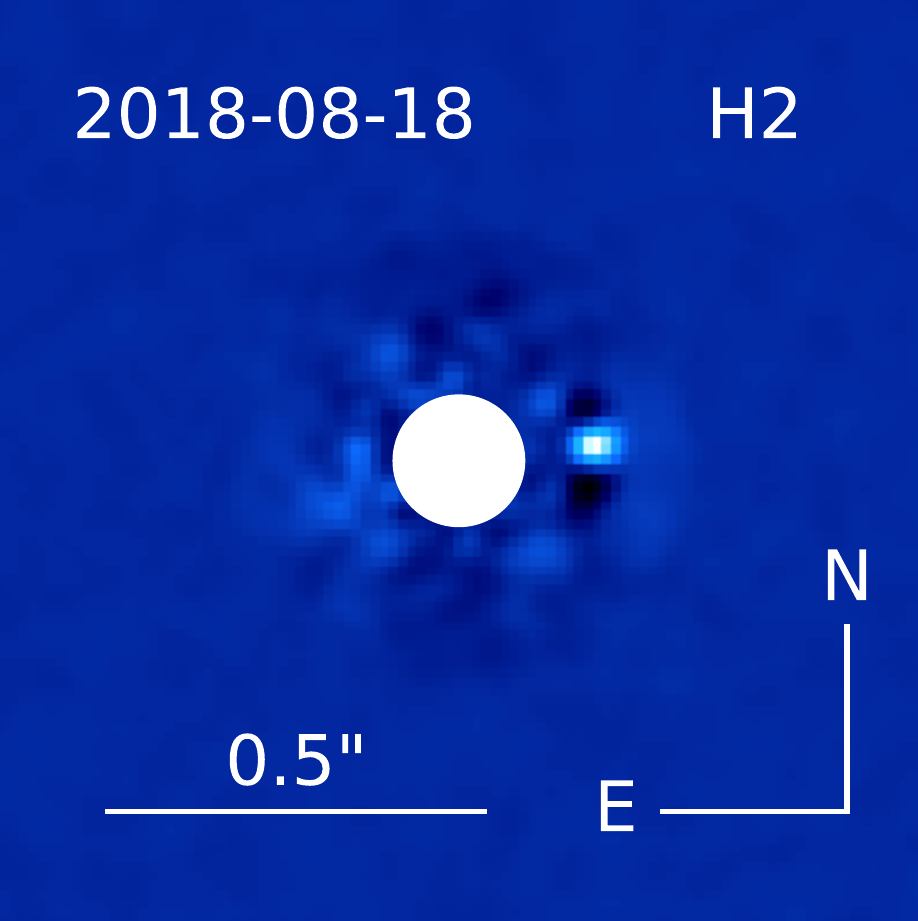}
\includegraphics[width=.28\textwidth]{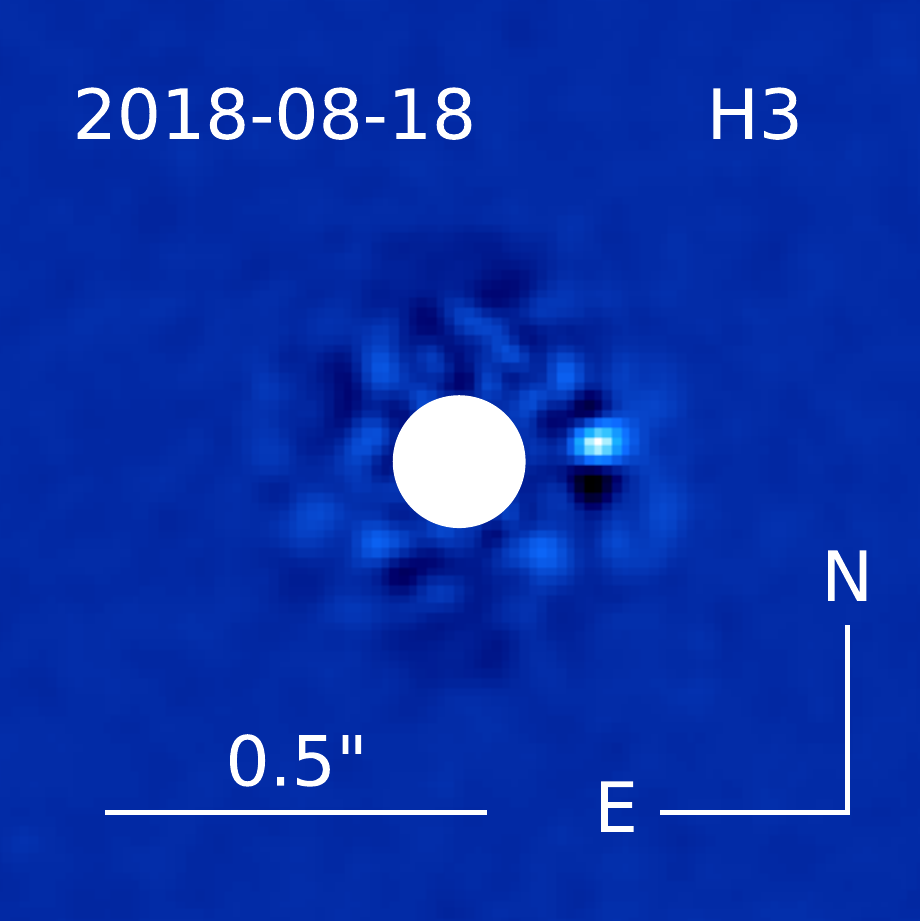}
\includegraphics[width=.28\textwidth]{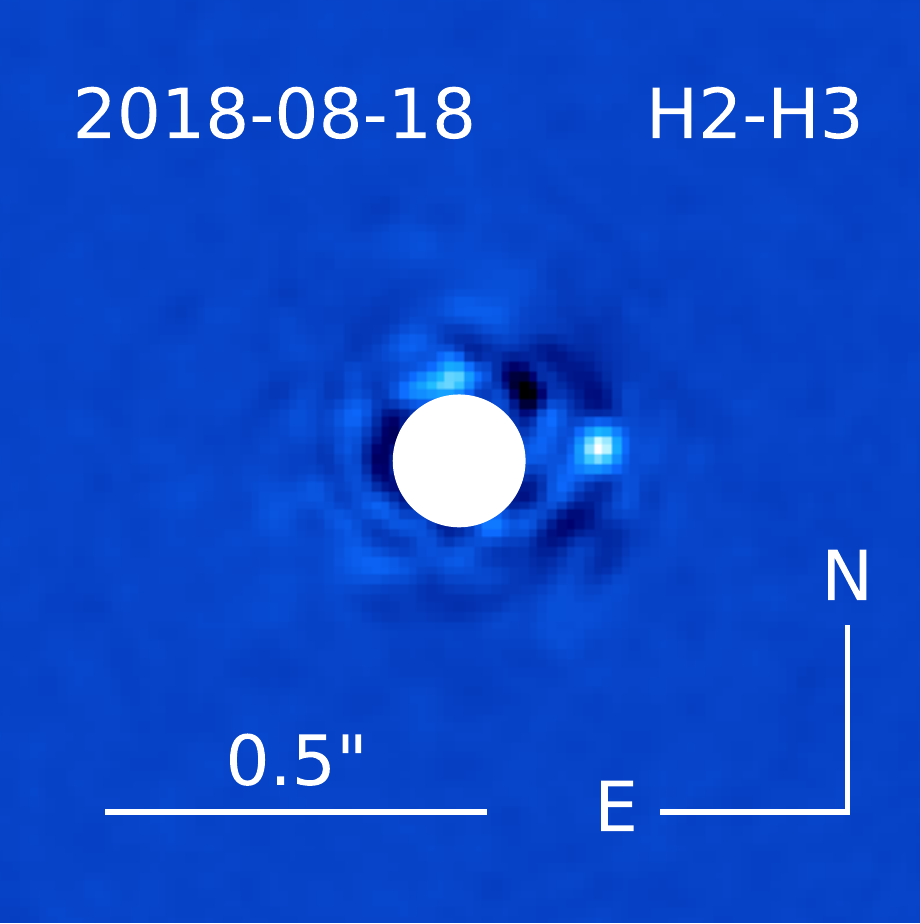}
\includegraphics[width=.28\textwidth]{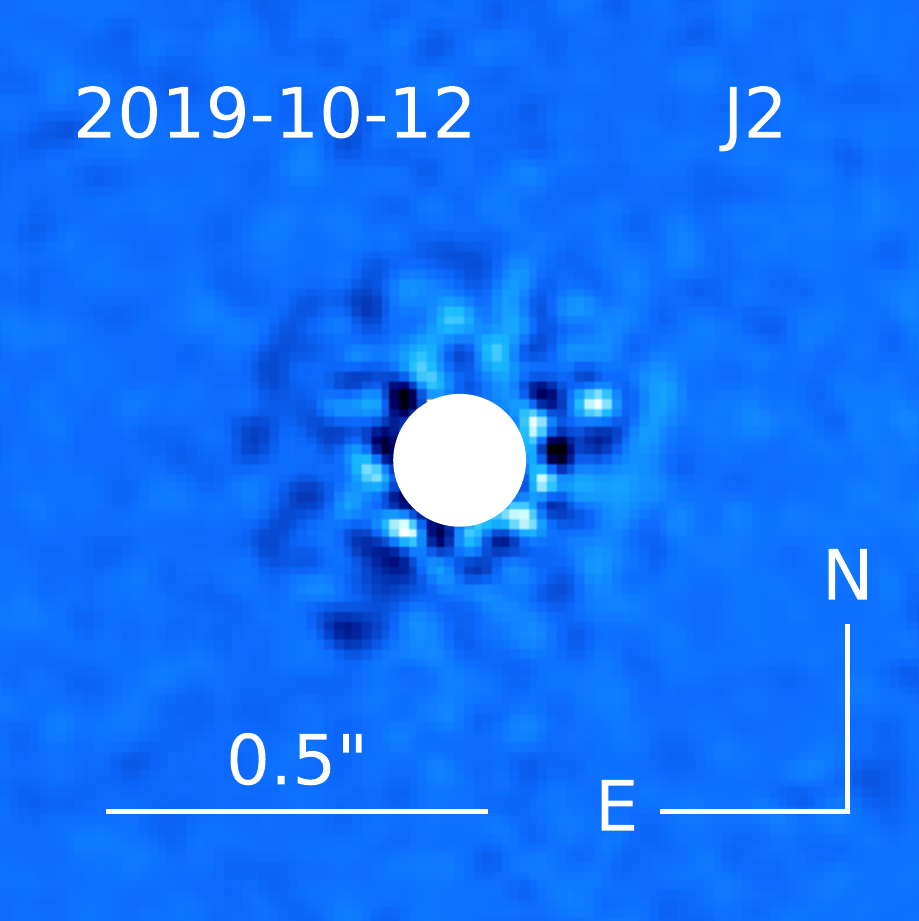}
\includegraphics[width=.28\textwidth]{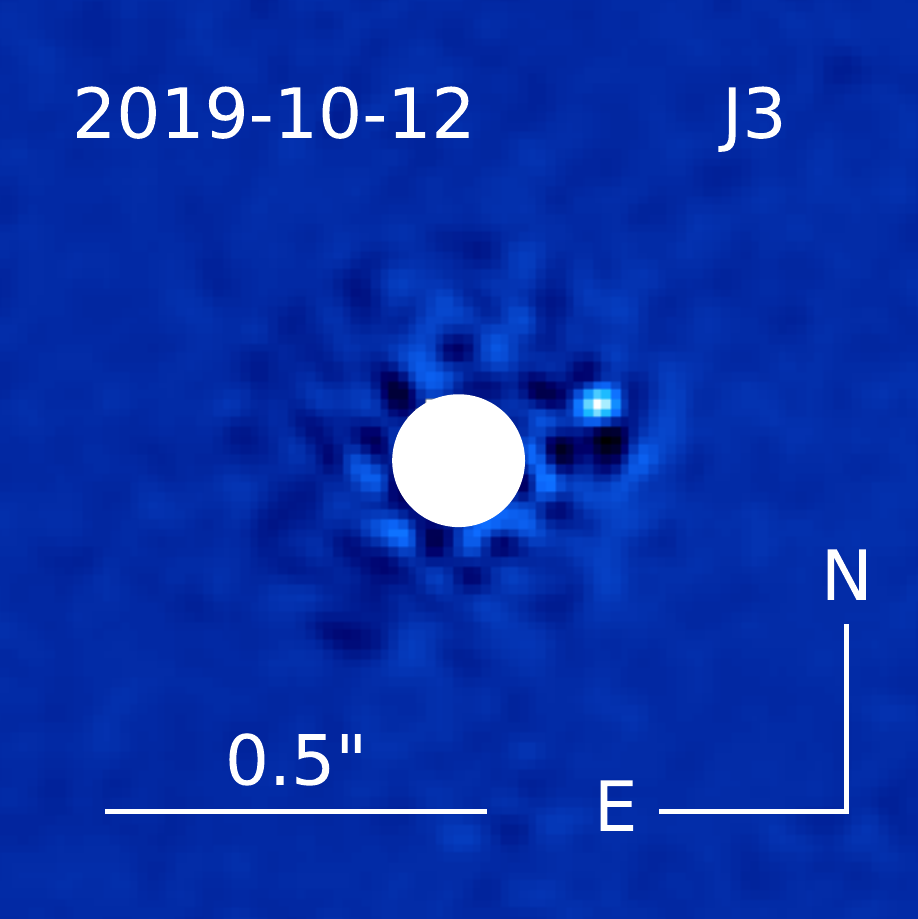}
\includegraphics[width=.28\textwidth]{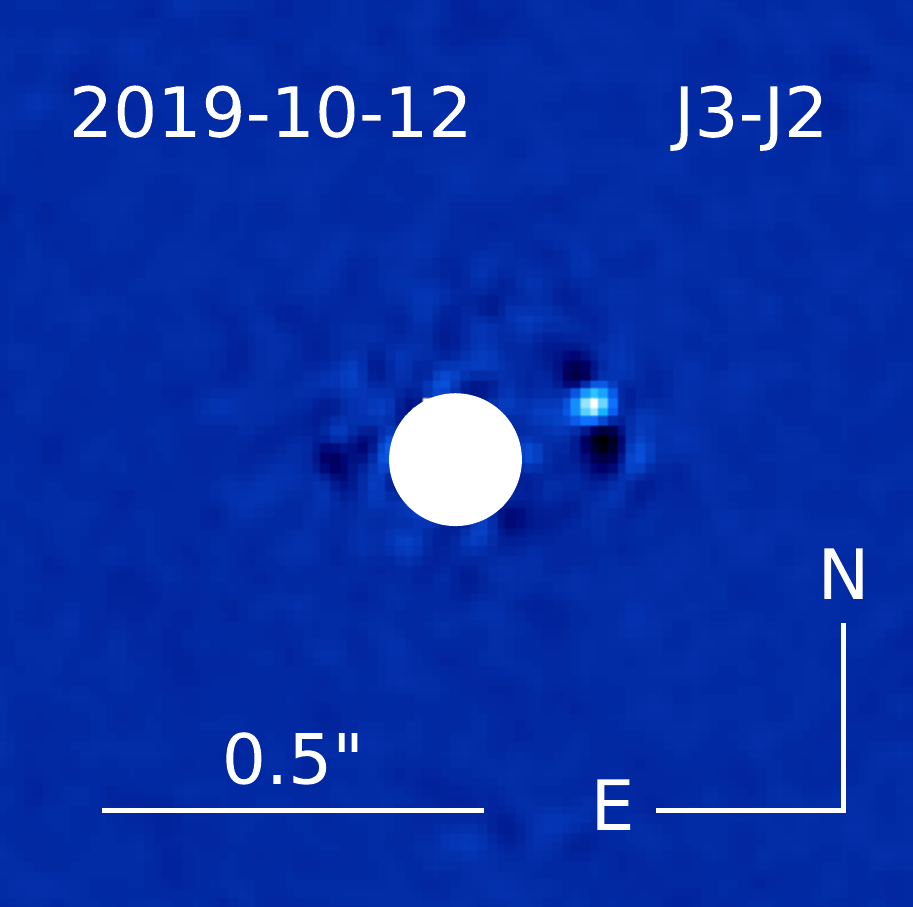}
\includegraphics[width=.28\textwidth]{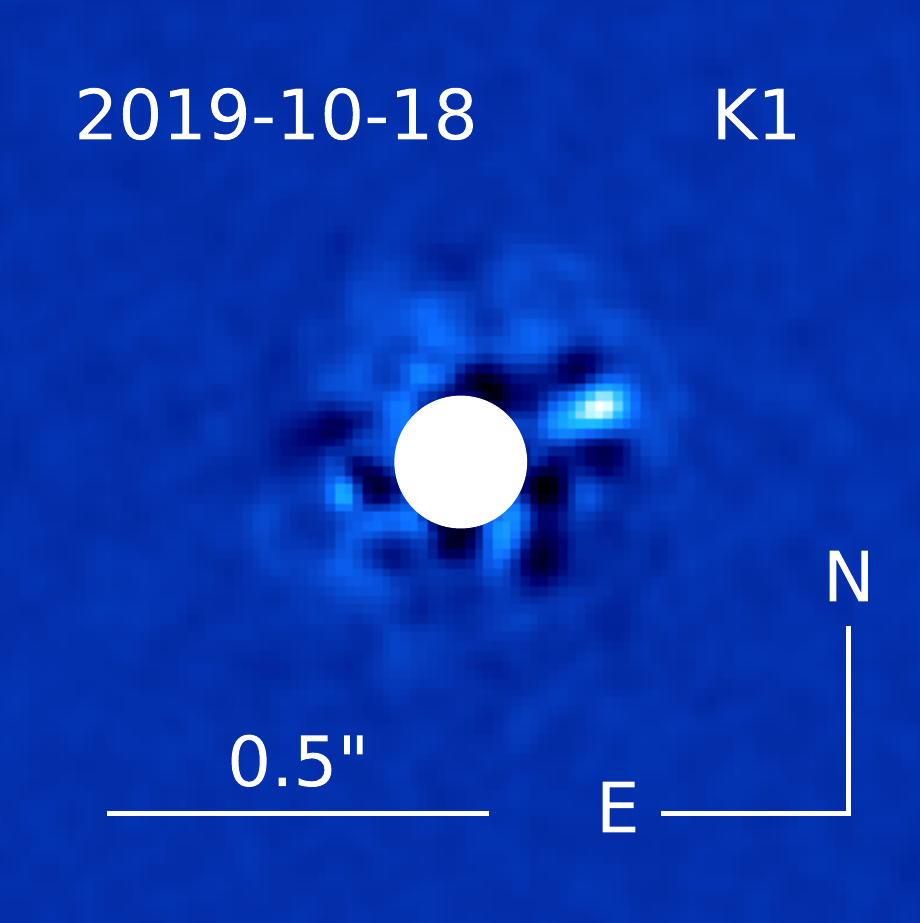}
\includegraphics[width=.28\textwidth]{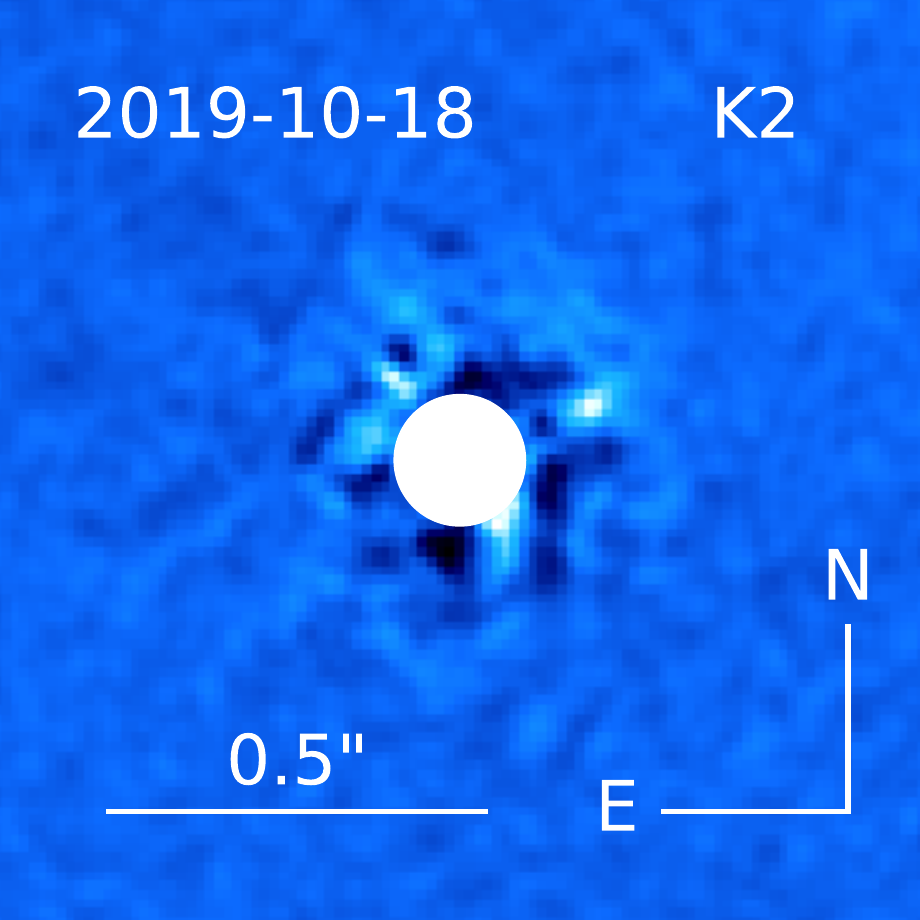}
\includegraphics[width=.28\textwidth]{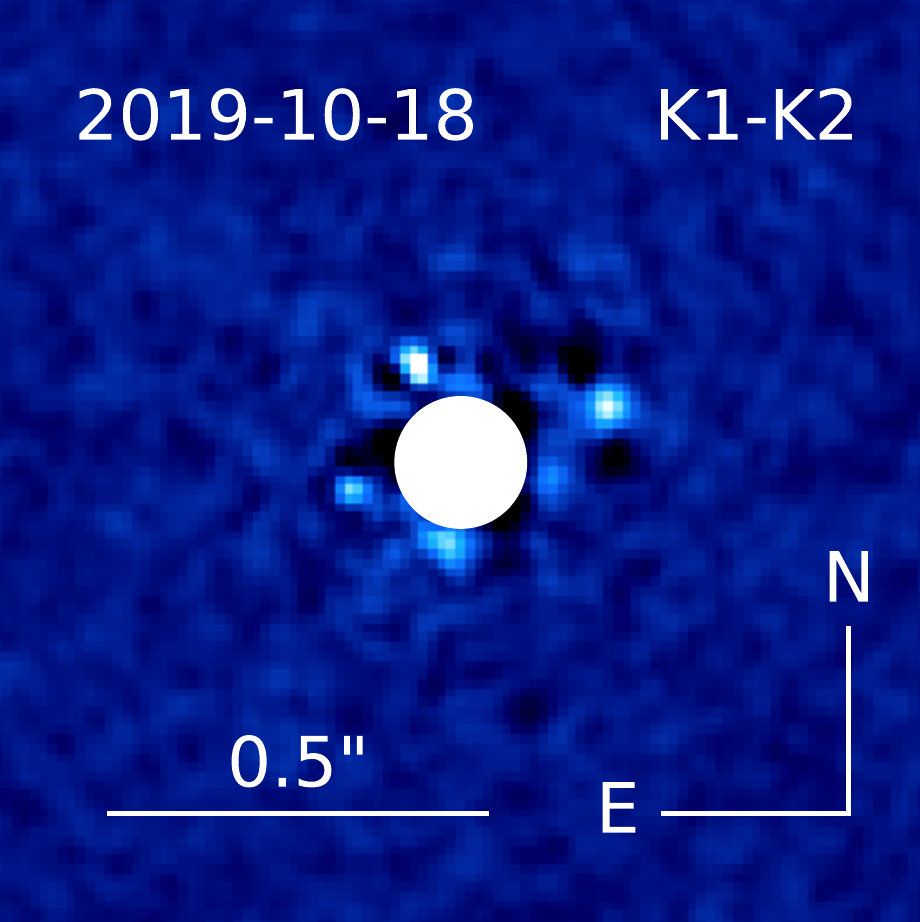}
\includegraphics[width=.28\textwidth]{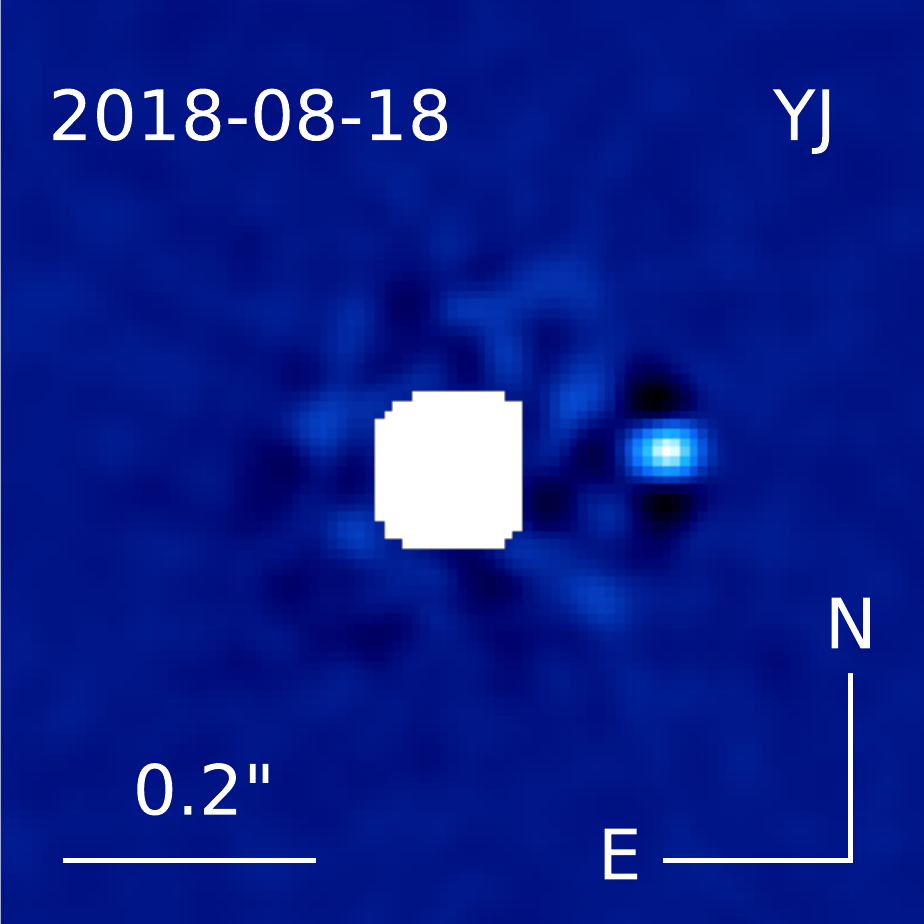}
\end{center}
\caption{IRDIS and IFS images of HD~13724. The star is masked by the white circle. We also show a weighted combination of the IFS SDI image. Each IFS wavelength channel was weighted by the average flux predicted by the best-fitting BT-Settl model \citep{2012RSPTA.370.2765A}.}
\label{fig:2}
\end{figure*}

HD~13724 was observed with SPHERE, the extreme adaptive optics system at the VLT \citep{2019arXiv190204080B} on 18 Aug 2018, 12 Oct 2019, and 18 Oct 2019. Observations were taken using the IRDIFS mode, which allows the Integral Field Spectrograph (IFS; \cite{2015A&A...576A.121M}) and the InfraRed Dual-Band Imager and Spectrograph (IRDIS; \cite{2008SPIE.7014E..3LD}) modules to be used simultaneously. The IFS data cover a range of wavelengths from $Y-J$ (0.96-1.34$\mu$m, spectral resolution $R \sim 54$). The IRDIS data were taken in  dual-band imaging mode \citep{2010MNRAS.407...71V} using the $H2$ and $H3$ filters ($\lambda_{H2} = 1.58888 \mu$m, $\lambda_{H3} = 1.6671 \mu$m), as well as using the $J2$ and $J3$ filters ($\lambda_{J2} = 1.190 \mu$m, $\lambda_{J3} = 1.273 \mu$m) and the $K1$ and $K2$ band filters ($\lambda_{K1} = 2.110 \mu$m and $\lambda_{K2} = 2.251 \mu$m), such that the $J2$, $H3,$ and $K2$ bands align with methane absorption bands. 

The observing sequence consisted of long-exposure images taken with an apodized Lyot coronagraph \citep{2003A&A...397.1161S}. To measure the position of the star behind the coronograph, several exposures were taken with a sinusoidal modulation applied to the deformable mirror (to generate satellite spots around the star) at the beginning and at the end of the sequence. To estimate the stellar flux and the shape of the point-spread function (PSF) during the sequence, several short-exposure images were taken with the star moved from behind the coronograph, and using a neutral-density (ND) filter with a $\sim10\%$ transmission\footnote{The corrections for the ND filter transmission are done using the ND filter curves given at \url{https://www.eso.org/sci/facilities/paranal/instruments/sphere/inst/filters.html}}, also at the beginning and end of the sequence. In addition, several long-exposure sky frames were taken to estimate the background flux and help identify bad pixels on the detector.

The SPHERE Data Reduction and Handling pipeline \citep{2008SPIE.7019E..39P} was used to perform the wavelength extraction for the IFS data, turning the full-frame images of the lenslet spectra into image cubes. The remainder of the data reduction and analysis was completed using the Geneva Reduction and Analysis Pipeline for High-contrast Imaging of planetary Companions (GRAPHIC) \citep{2016MNRAS.455.2178H}. The data were first sky subtracted, flat fielded, cleaned of bad pixels, and corrected for distortion (following \citet{2016SPIE.9908E..34M}). We then ran a principal component analysis (PCA) PSF substraction algorithm \citep{2012ApJ...755L..28S, 2012MNRAS.427..948A} which is run separately for images at each wavelength channel and for each IRDIS channel. The resulting frames were derotated and median combined to produce a final PSF-substracted image. 

In addition to this, we performed a spectral differential imaging (SDI) reduction for both the IFS and IRDIS datasets. The same PCA algorithm was then performed on the resulting images. The resulting PCA-reduced and SDI IRDIS and IFS images are shown in Fig.~\ref{fig:2}. 

\section{Results}

\subsection{Astrometry and photometry}

The relative astrometry and photometry were calculated using a negative fake-planet injection on the images \citep{2011A&A...528L..15B}. An initial Nelder-Mead minimisation routine \citep{Gao2012} was used to find a first-order estimation of the separation (in pixels), position angle (in degrees), and contrast ratio between the star and the companion. We then fitted over a 3D grid of parameters for the position and the flux of the fake negative PSF to minimise the residuals and to numerically derive the $\chi^2$ with the associated 68.27\% confidence interval. Subsequently, we corrected for the pixel scale and true north angle given in \citet{2016SPIE.9908E..34M}. The resulting parameters are listed in Table \ref{tab:3}.

The IFS data were taken in the wavelength range 0.96-1.34$\mu$m which is split into 39 wavelength channels, giving 39 images across the wavelength range. Because of the low signal-to-noise ratio in some of the spectral channels, we used a different approach to measure the companion flux. For the IFS data, we performed a fit for the separation and position angle using the negative fake-planet injection using a stacked image across all of the wavelength channels in order to have a high signal-to-noise ratio for the companion fit. Using this fitted position on the image, we then fitted for the contrast ratio between the companion and the star at each wavelength channel in order to extract the spectrum.

\begin{table*}
    \caption{Measured relative astrometry and photometry of HD~13724~B from the SPHERE observations.}
    \centering
    \begin{tabular}{ccccccc}
    \hline
    \hline
    Instrument & Filter & Date & BJD \tablefoottext{a} & $\rho$ (mas) & $\theta$ (deg) & Contrast (mag) \\
    \hline
         IRDIS & H2 & 2018-08-18 &58349.29438& 175.61 $\pm$ 4.45 & 272.15 $\pm$ 1.06 & 10.61 $\pm$ 0.16 \\
         IRDIS & H3 & 2018-08-18 &58349.29438& 178.85 $\pm$ 4.56 & 272.50 $\pm$ 1.71 & 11.34 $\pm$ 0.32 \\
         IRDIS & H2-H3 & 2018-08-18 &58349.29438& 177.70 $\pm$ 4.41 & 270.75 $\pm$ 1.76 & 11.71 $\pm$ 0.35 \\
         IRDIS & J2 & 2019-10-12 &58768.16564& 179.9 $\pm$ 12.69 & 289.92 $\pm$ 2.49 & 11.50 $\pm$ 0.40 \\
         IRDIS & J3 & 2019-10-12 &58768.16564& 188.90 $\pm$ 4.98 & 288.75 $\pm$ 1.05 & 10.37 $\pm$ 0.04 \\
         IRDIS & J3-J2 & 2019-10-12 &58768.16564& 187.59 $\pm$ 1.60 & 288.25 $\pm$ 0.58 & 10.82 $\pm$ 0.09 \\
         IRDIS & K1 & 2019-10-18 &58774.20206& 190.55 $\pm$ 9.05 & 288.75 $\pm$ 1.51 & 10.29 $\pm$ 0.18 \\   
         IRDIS & K2 & 2019-10-18 &58774.20206& 192.12 $\pm$ 15.84 & 288.75 $\pm$ 2.52 & 11.10 $\pm$ 0.46 \\
         IRDIS & K1-K2 & 2019-10-18 &58774.20206& 200.32 $\pm$ 16.15 & 288.58 $\pm$ 2.62 & 11.15 $\pm$ 0.43 \\
    \end{tabular}
\tablefoot{\tablefoottext{a}{The date are expressed as BJD-2400000.0.}}
    \label{tab:3}
\end{table*}

\subsection{Orbit determination and dynamical mass}

To constrain the orbital parameters of the brown dwarf, we performed a combined fit to the RV and direct-imaging data. For this we used the IRDIS $H2$, $H3$, $J2$, $J3$, $K1$ and $K2$ astrometry $\{ \rho(t), \theta(t) \}$ where $\rho(t)$ is the observed separation in mas and $\theta(t)$ is the position angle in degrees (see Table~\ref{tab:4}). This allows us to place constraints on the period, eccentricity, and inclination of the system.

Due to two major upgrades to CORALIE in June 2007 \citep{2010A&A...511A..45S} and in November 2014, we consider CORALIE as three different instruments, corresponding to the different upgrades: the original CORALIE as CORALIE-98 (C98), the first upgrade as CORALIE-07 (C07), and the latest upgrade as CORALIE-14 (C14).

The observed RV signal was modelled with a single Keplerian and five RV offsets (corresponding to one offset for each RV instrument). \textsc{HARPS-03} was chosen as the reference instrument while RV offsets were adjusted between \textsc{CORAVEL}, \textsc{CORALIE-98}, \textsc{CORALIE-07,} \textsc{CORALIE-14,} and \textsc{HARPS-03;} where each of the instrumental offsets were margnialised over for the fit.

A linear correlation is observed between the $H_\alpha$ activity index time series and the observed RVs which allows us to carry out a first-order detrending of the observed RV measurements using a linear scale factor between the $H_\alpha$ index time series and the modelled RVs \citep{2018A&A...614A.133D}. A RV nuisance parameter -- corresponding to the white noise component of the stellar activity -- is added to the noise model of the likelihood function and is adjusted in the MCMC.

Regarding the Keplerian motion, we choose to adjust the natural log of the period and of the RV semi-amplitude to increase the efficiency of the MCMC due to the partial coverage of the orbit. We also probe the eccentricity and the argument of periastron through $\sqrt{e}.\cos{\omega}$ and  $\sqrt{e}.\sin{\omega}$ variables. We use a uniform prior on these variables which also corresponds to a uniform prior in eccentricity and $\omega$. The longitude of the ascending node, $\Omega$, the inclination, $i$, and the relative orbit semi-major axis expressed in milli-arcseconds are also adjusted. The conversion from angles to astronomical units is done using the \emph{Gaia} parallax as a Gaussian prior. The orbit phase reference was chosen as the time at which the RV is minimum, $T_{\rm Vmin}$, since it is well defined by the RV observations (see Fig.\ref{fig:1}).

We probe the full parameter space, composed of 16 parameters, using an MCMC with an adaptive Metropolis \citep{haario2001} and an adaptive scaling \citep{andrieu2008} which is particularly efficient at probing parameters with linear correlations. Additional tables and figures illustrating the results of the MCMC analysis are provided in the Appendix.

\begin{figure}
\includegraphics[width=0.5\textwidth]{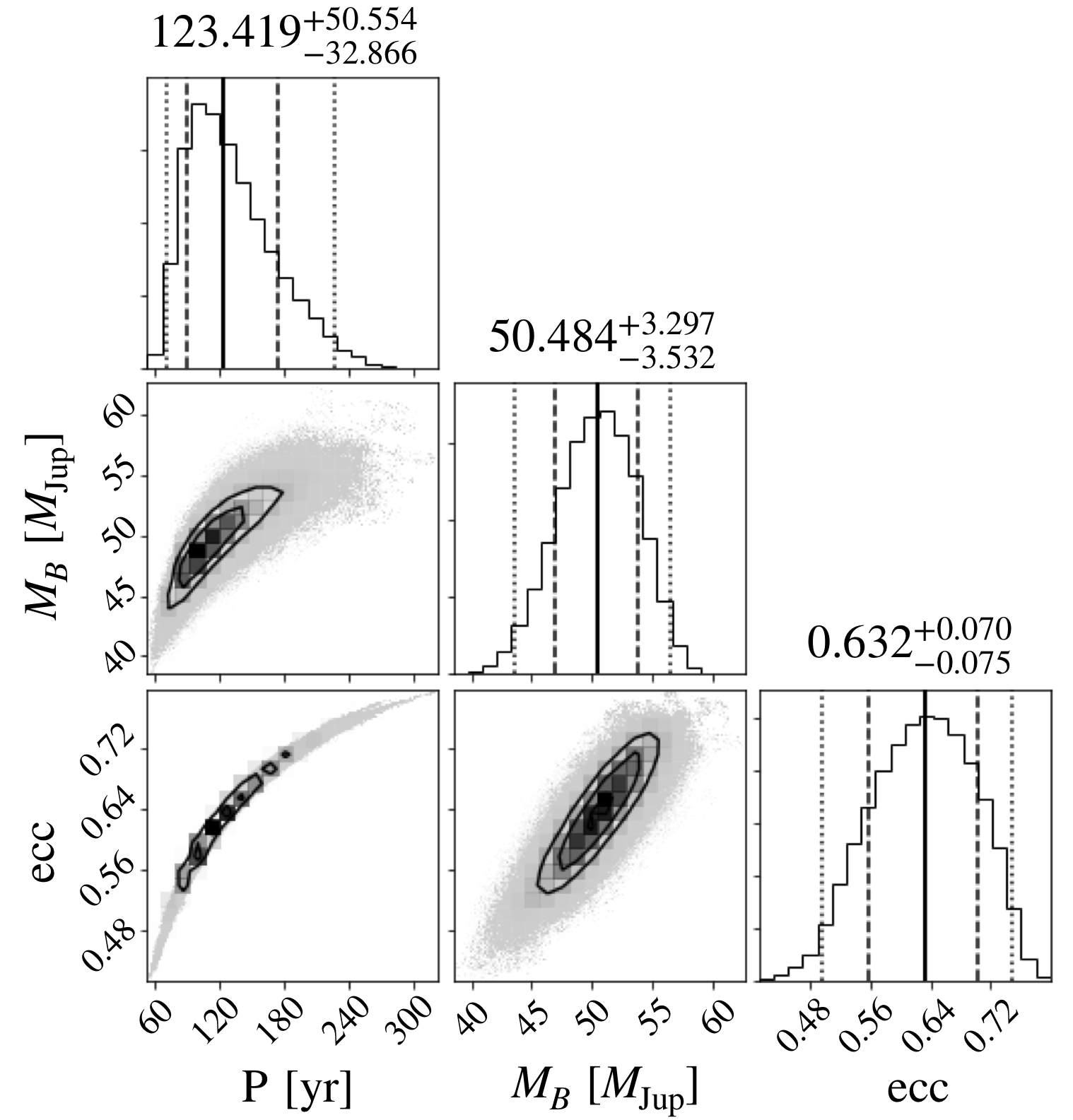}
\caption{Marginalised 1D and 2D posterior distributions of the orbital period, as well as the eccentricity and the companion mass corresponding to the global fit of the RV and direct-imaging models. Confidence intervals at 2.275\%, 15.85\%, 50.0\%, 84.15\%, 97.725\% are overplotted on the 1D posterior distributions, while the median $\pm 1~\sigma$ values are given at the top of each 1D distribution. Here, 1, 2, and 3~$\sigma$ contour levels are overplotted on the 2D posterior distribution.}
\label{fig:3}
\end{figure}

From combining the RV and direct imaging measurements we were able to bring good constraints on the geometry of the orbit and the mass of HD~13724~B. Based on the MCMC posterior distribution, we are able to set confidence intervals for the orbital elements and physical parameters of HD~13724~B. The 95\% confidence interval for the period ranges from 72 to 226~yr, its semi-major axis between 18 and 40 AU, and its mass between 43 and 56~$M_{\mathrm{Jup}}$. Orbital elements values and confidence intervals are given in Table~\ref{tab:ds1} while the full list of parameters adjusted in the MCMC is provided in the Appendix (Table~\ref{tab:ds2}). A mass-period-eccentricity corner plot based on the marginalised 2D posterior distribution is presented in Fig.~\ref{fig:3} and illustrates the wide range of orbits still compatible with our data. However, the nature of HD~13724~B is undoubtedly well below the hydrogen burning limit.

We note that these orbital solutions have higher eccentricities and longer periods than those published in \cite{2019A&A...625A..71R}. This is explained by (1) the SPHERE measurements that rule out the shortest periods and (2) a significant improvement in our MCMC implementation that increases the efficiency with which we can probe correlated parameter spaces (see Figure~\ref{fig:A1}).

\begin{table}
    \caption{Orbital elements of HD13724~B derived from the MCMC posterior distributions}
    \centering
    \begin{tabular}{cccccc}
    \hline
    \hline
    Param & Unit & Med & Std & CI(2.5\%) & CI(97.5\%) \\
    \hline
         $P$&yr&123&41&72&226\\
         $K$&m/s&278&13&251&299\\
         e&&0.63&0.07&0.5&0.75\\
         $\omega$&deg&184.9&1.1&182.8&187.0\\
         $\Omega$&deg&16.4&4.2&7.3&23.9\\
         $i$&deg&57.4&1.7&53.6&60.9\\
         $a_r$&AU&26.3 & 5.6 & 18.3 & 39.5\\
         $T_{\rm Vmin}$&d\tablefoottext{a}&56034&21&55993&56074\\
         \hline
         $M_{\rm B}$ & M$_{\rm Jup}$ & 50.5&3.3&43.7&56.7\\
         \hline
    \end{tabular}
\tablefoot{\tablefoottext{a}{The date is expressed as BJD-2400000.0.}}
    \label{tab:ds1}
\end{table}

\subsection{Companion properties}

To allow a consistent comparison with other objects and to estimate the mass of HD~13724~B, we converted the $H2$, $H3$, $J2$, $J3$, $K1$ and $K2$ fluxes into absolute magnitudes using the \emph{Gaia}-measured distance of HD~13724 \citep{2018A&A...616A...1G} shown in Table~\ref{table:1}. We then used the COND \citep{2003A&A...402..701B} substellar isochrones to predict the companion mass and temperature as a function of the calculated absolute photometry in each band. The resulting absolute photometric magnitudes, masses, and temperatures are listed in Table~\ref{tab:4}.

\begin{table}
    \caption{The measured absolute photometry of HD~13724~B with the derived mass and temperature using \citet{2003A&A...402..701B}.}
    \centering
    \footnotesize
    \begin{tabular}{ccccc}
    \hline
    \hline
         Band & App. Mag & Abs. Mag & Mass & Temp. \\
          &  &  & ($M_{\mathrm{Jup}}$) & (K) \\
         \hline
         H2 & 17.09 $\pm$ 0.16 & 13.90 $\pm$ 0.16 & 43.9 $\pm$ 1.8 & 1306 $\pm$ 48 \\ 
         H3 & 17.82 $\pm$ 0.32 & 14.63 $\pm$ 0.32 & 36.7 $\pm$ 3.0 & 1128 $\pm$ 74 \\
         J2 & 18.23 $\pm$ 0.40 & 15.04 $\pm$ 0.40 & 32.4 $\pm$ 4.0 & 1020 $\pm$ 99 \\
         J3 & 17.10 $\pm$ 0.05 & 13.91 $\pm$ 0.05 & 44.0 $\pm$ 0.6 & 1310 $\pm$ 18 \\
         K1 & 16.67 $\pm$ 0.18 & 13.48 $\pm$ 0.18 & 47.5 $\pm$ 2.2 & 1399 $\pm$ 56 \\
         K2 & 17.48 $\pm$ 0.46 & 14.29 $\pm$ 0.46 & 39.0 $\pm$ 4.3 & 1184 $\pm$ 106 \\
    \hline
    \end{tabular}
    \label{tab:4} 
\end{table}

\subsection{Spectral and atmospheric analysis}

In order to estimate the spectral type of HD~13724~B, we used the SpeX Prism library of near-infrared (NIR) spectra of brown dwarfs \citep{2014ASInC..11....7B} using the \textit{splat} python package \citep{2016AAS...22743408B} where each spectrum was flux calibrated to the distance of HD~13724. The \textit{splat} package contains a library of observed brown dwarf spectra as well as theoretical models that we use as templates to derive the physical parameters.

Each SpeX Prism spectrum was also converted into the appropriate spectral resolution of each IFS measurement by convolution with a Gaussian. The FWHM used for the convolution was assumed to be twice the separation between each wavelength channel. To fit the spectrophotometry of HD~13724~B with atmospheric models, we converted the contrast measurements into physical fluxes using a model spectrum for the host star ($T_{\mathrm{eff}}$ = 5900~K, $\log g$ = 4.5~dex and [Fe/H] = 0.3~dex) from the BT-NextGen library \citep{2012RSPTA.370.2765A} and the SPHERE filter transmission curves. The BT-NextGen spectrum is fit to the spectral energy distribution (SED) and is built using data from Tycho \citep{2000A&A...355L..27H}, 2MASS \citep{2003yCat.2246....0C}, WISE \citep{2013yCat.2328....0C}, and HIPPARCOS \citep{1997A&A...323L..49P}.

\begin{figure*}
\centering
\includegraphics[width=0.97\textwidth]{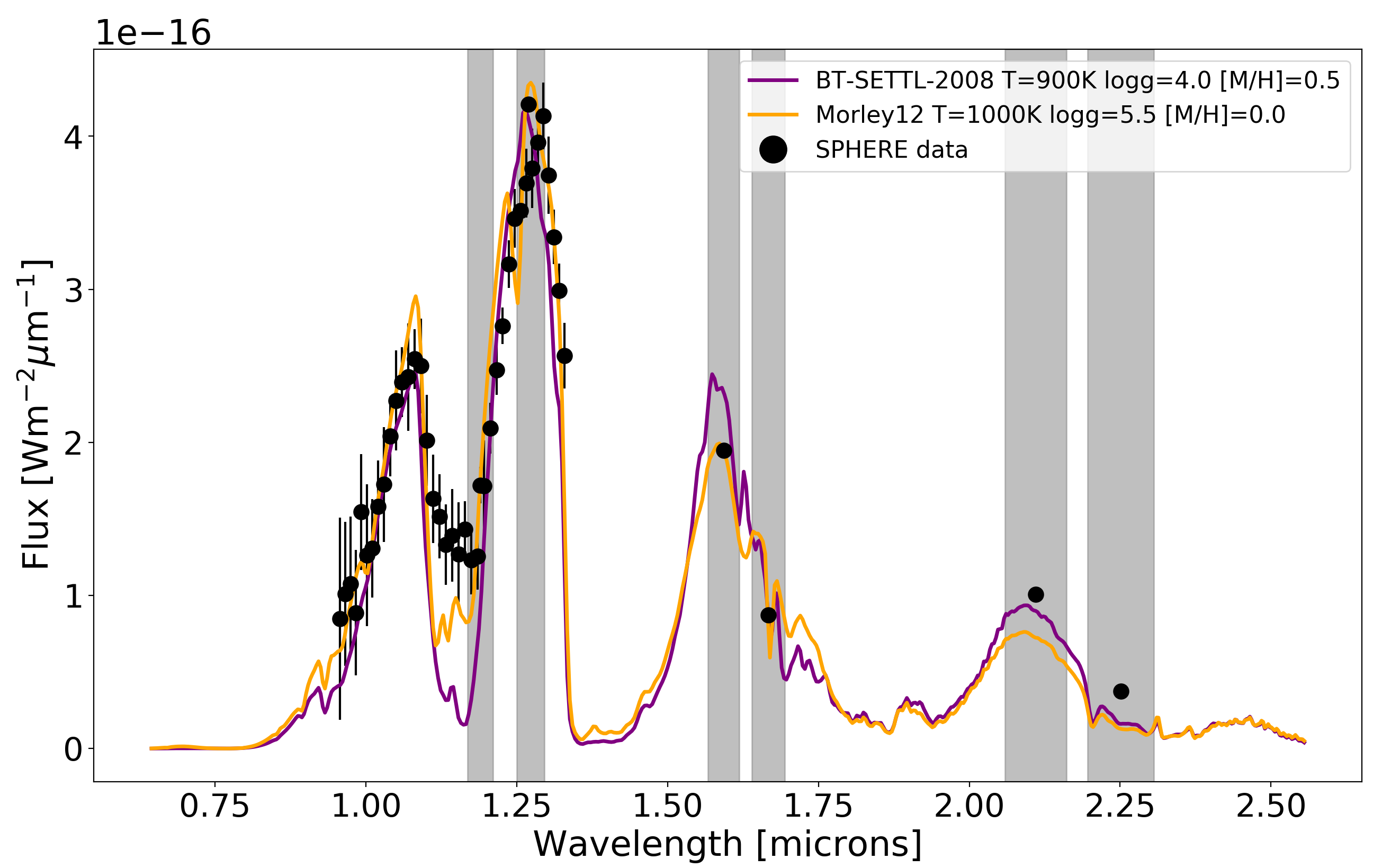}
\includegraphics[width=0.99\textwidth]{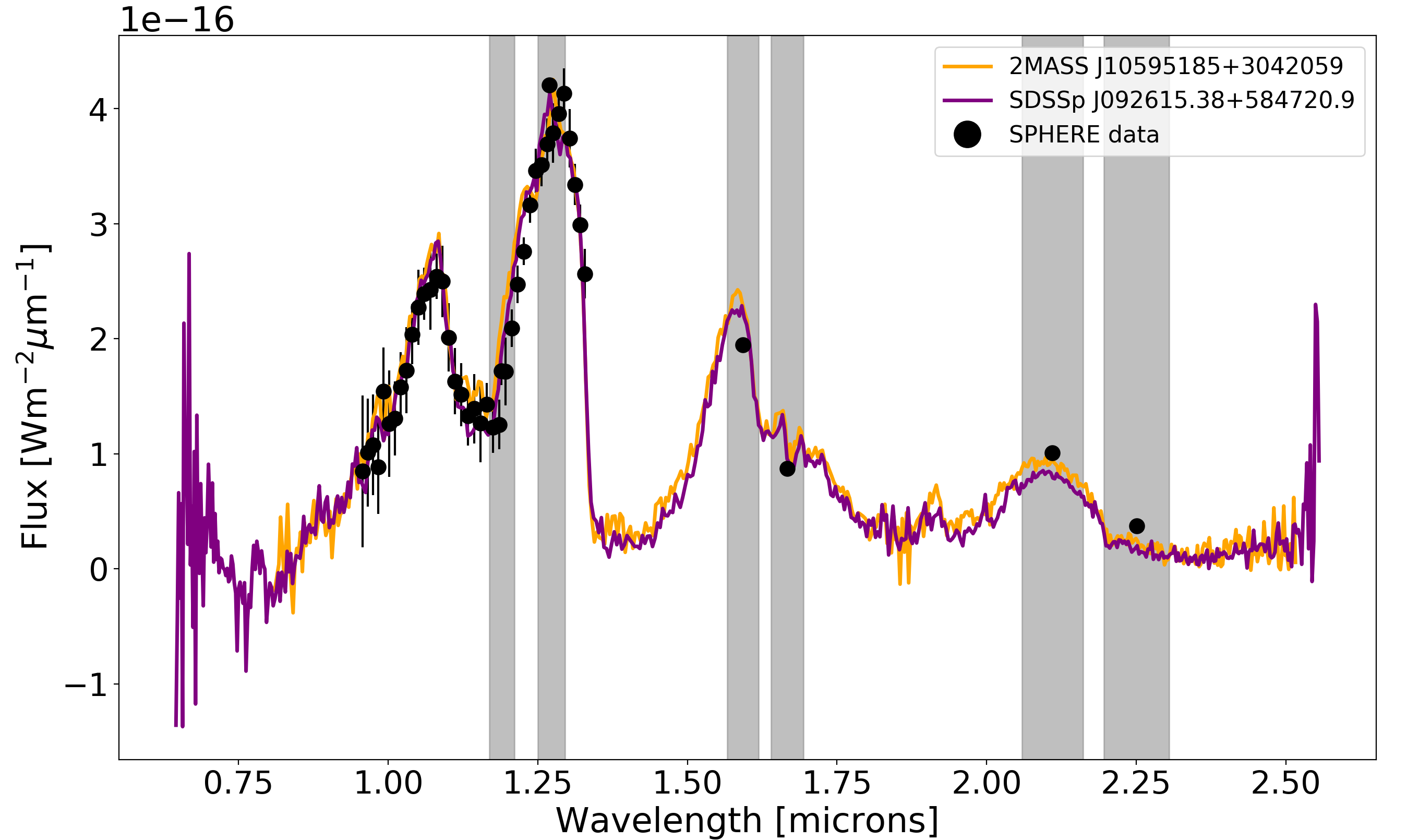}
\caption{Atmospheric spectra of HD~13724. The FWHM of the $J23$, $H23,$ and $K12$ IRDIS filters are shown with grey bars (from left to right). The error bars on the $H2$, $H3$, $K1,$ and $K2$ data are present but smaller than the size of the markers on each plot. \textbf{Top:} The two best-fitting spectra from BT-Settl 2008 models \citep{2012RSPTA.370.2765A} and the \cite{2012ApJ...756..172M} models from the SpeX Prism library are overplotted, and the flux is scaled to match HD~13724~B. Fitting was performed to the IFS and IRDIS data simultaneously. \textbf{Bottom:} Comparison of the observed spectrum of HD~13724~B with the two best fit T4 brown dwarf atmospheres: 2MASS J10595185+3042059 and SDSSp J092615.38+58720.9 fitted using the \emph{splat} package \citep{2016AAS...22743408B}.}
\label{fig:4}
\end{figure*}

We calculated the $\chi_{r}^2$ of template brown dwarfs in the SpeX Prism library as a function of spectral type. We include the uncertainties of the template spectra in the $\chi^2$ computation. The best-fit object is 2MASS J10595185+3042059 \citep{2003yCat.2246....0C} which is classified as a T4 brown dwarf. In Fig.~\ref{fig:4} we also plot the second-best fit spectrum of SDSSp J092615.38+584720.9 \citep{2003yCat.2246....0C}, which is classified as a T4.0/T4.5 brown dwarf. The temperatures of 2MASS J10595185+3042059 and SDSSp J092615.38+584720.9 are not given explicitly but effective temperatures of mid-T dwarfs covers a small range ($T_{\mathrm{eff}} \approx 1100-1400$~K \cite{2000AJ....120..447K}). The T4 spectral type provides the best fit.

\begin{figure}
\includegraphics[width=0.5\textwidth]{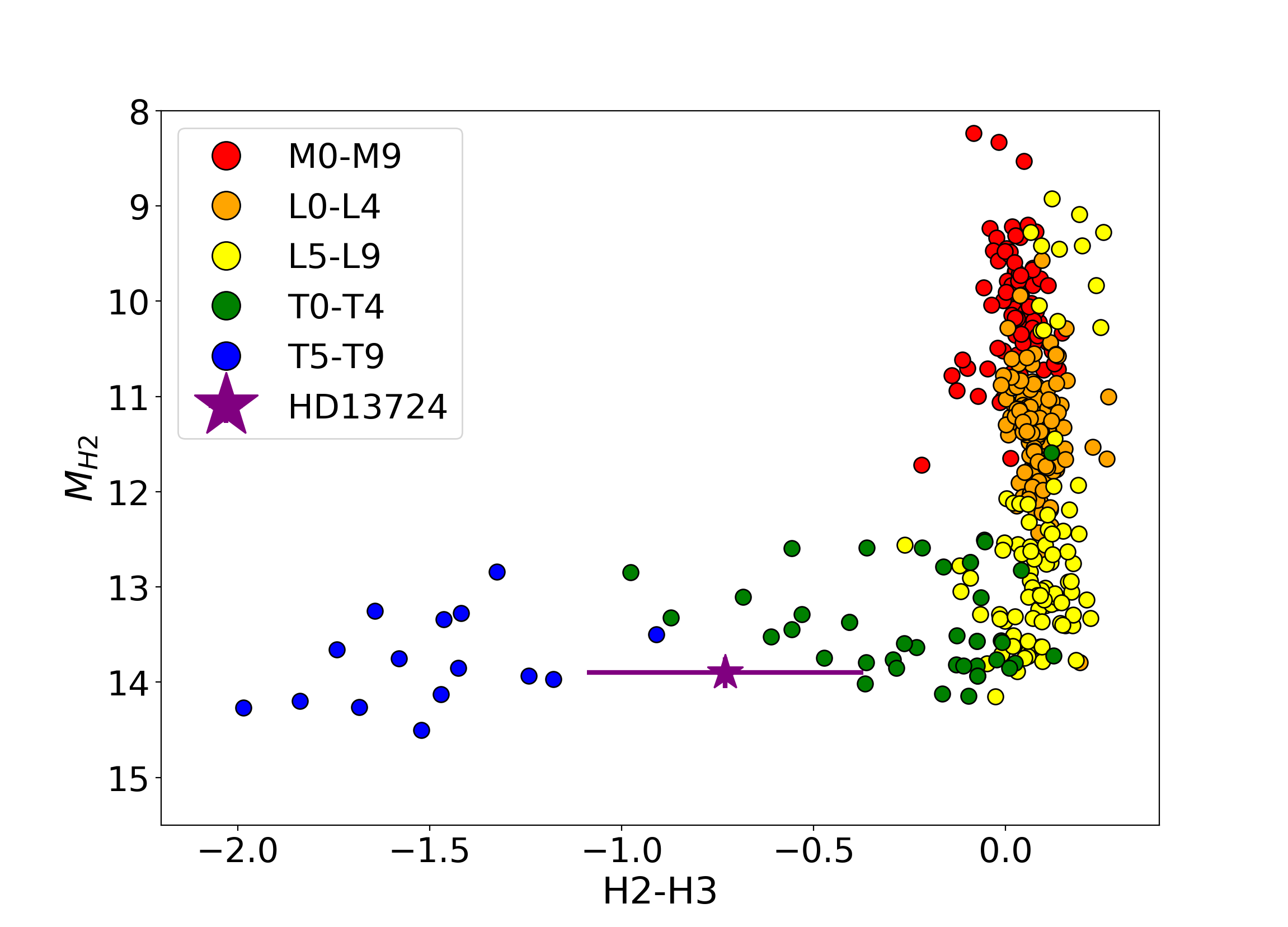}
\caption{Colour-magnitude diagram showing the predicted H2-H3 colours
and H2 absolute magnitudes for objects in the SpeX Prism Library \citep{2014ASInC..11....7B}. The flux of HD~13724~B and the 1$\sigma$ limit on its H2-H3 colour are shown. Its position is compatible with objects in the middle of the T sequence.}
\label{fig:5}
\end{figure}

We use the $H2$ and $H3$ band values to place HD~13724~B on a colour-magnitude diagram as shown in Fig.~\ref{fig:5} which is built using the Spex Prism Library \citep{2014ASInC..11....7B}. The placement of HD~13724~B on the colour-magnitude diagram agrees well with the estimated T spectral type.

To further constrain the physical properties of HD~13724~B, we compared the observed SED to synthetic spectra for cool brown dwarfs from \cite{2012ApJ...756..172M} and \cite{2012RSPTA.370.2765A}. The biggest difference between these models is that the \cite{2012ApJ...756..172M} models include the condensate sedimentation efficiency coefficients.

We adjusted the \cite{2012ApJ...756..172M} and \cite{2012RSPTA.370.2765A} models, which sit well within the range of the mid-T dwarf range (see Fig.~\ref{fig:4}). The  \cite{2012ApJ...756..172M} models are the best fit to the data with the following parameters: $T_{\mathrm{eff}}$ = 1000~K, $\log g = 5.5 \mathrm{cm}/\mathrm{s}^2$, $f_{\mathrm{SED}}$ = 3, [M/H] = 0.0, and $R=1.01~R_{\mathrm{Jup}}$.

\section{Summary and conclusions}

In this paper we report the direct detection of a $\sim$ 50 $M_{\mathrm{Jup}}$ brown dwarf using VLT/SPHERE. HD~13724~B serves as an essential benchmark brown dwarf for testing brown dwarf atmospheric and evolutionary models where it joins just a short list of benchmark brown dwarf companions of stars with RV and imaging measurements: HR~7672~B \citep{2002ApJ...571..519L, 2012ApJ...751...97C}, HD~19467~B \citep{2014ApJ...781...29C}, HD~4747~B \citep{2011A&A...525A..95S, 2016ApJ...831..136C, 2019A&A...631A.107P}, GJ~758~B \citep{2009ApJ...707L.123T, 2018AJ....155..159B}, HD~4113~C \citep{2018A&A...614A..16C}, GJ~229~B \citep{1995Natur.378..463N, 2019arXiv191001652B, 2020arXiv200102577F}, and HD~72946~B \citep{2019arXiv191202565M}.

We obtained images of HD~13724~B in the \textit{J}, \textit{H,} and \textit{K} bands along with its spectrum from IFS data in the $Y-J$ bands. Through comparison with the SpeX Prism Library of brown dwarf spectra \citep{2014ASInC..11....7B}, we found that HD~13724~B is most consistent with a spectral type of T4, which also agrees with the position of HD~13724~B on the $H2-H3$ colour-magnitude diagram (Fig.~\ref{fig:5}).
It should be noted that in general the template T4 brown dwarf provides a better fit than the synthetic spectra of \cite{2012ApJ...756..172M}. This is especially true in the $K1$ and $K2$ bands where we observe more flux than the models would suggest. This might be due to clouds in the atmospheres that tend to increase the flux in the $K$-band \citep{1996Sci...272.1919M, 2012ApJ...756..172M}. The template adjustment using the \cite{2012ApJ...756..172M} models allowed us to derive a temperature of $1000~K$. 

We analysed the HD~13724 system by combining RV measurements and high-contrast imaging, enabling us to place constraints on the orbital and physical parameters of HD~13724~B. This suggests a mass of 43-56~$M_{\mathrm{Jup}}$, an inclination of 53-60~deg, and a period range of 72-226 years.
We calculated the isochronal mass of HD~13724~B from the COND substellar isochrones \citep{2003A&A...402..701B} taking into account the age of the star. This gives a mass range of 31-52~$M_{\mathrm{Jup}}$ corresponding to an age of 0.5-2 Gyr. This overlaps with the mass range from the observations and suggests that either the star is slightly older than expected or -- if the
age is correct -- some physics is missing from the modelling of the evolution of these substellar objects.

The combination of \emph{Gaia} astrometry with future RV and imaging datasets will allow for much tighter constraints on the orbital and physical parameters of HD~13724~B, which in turn will improve the comparison with atmospheric and evolutionary models.

Many attempts have been made to detect these long-period companions from the CORALIE RV survey through imaging as part of a fifteen-year effort using VLT/NACO, with little success. The VLT/SPHERE is allowing us to now unveil these RV-detected brown dwarfs. We expect that the future upgrade of SPHERE \citep[SPHERE+][]{2019vltt.confE..38B} will increase the sample of cool brown dwarfs detected thanks to its higher contrast capabilities at short separations. Ultimately, future instruments like SPHERE+, the James Webb Space Telescope, and the upcoming ELT/METIS \citep{2016SPIE.9909E..73C} should allow us to bridge the gap between the coolest brown dwarfs and the most massive giant planets.

\begin{acknowledgements}
This work has been carried out within the framework of the National Centre for Competence in Research PlanetS supported by the Swiss National Science Foundation. The authors acknowledge the financial support of the SNSF.
This publications makes use of the The Data \& Analysis Center for Exoplanets (DACE), which is a facility based at the University of Geneva (CH) dedicated to extrasolar planets data visualisation, exchange and analysis. DACE is a platform of the Swiss National Centre of Competence in Research (NCCR) PlanetS, federating the Swiss expertise in Exoplanet research. The DACE platform is available at \url{https://dace.unige.ch}. This work has made use of data from the European Space Agency (ESA) mission {\it Gaia} (\url{https://www.cosmos.esa.int/gaia}), processed by the {\it Gaia} Data Processing and Analysis Consortium (DPAC, \url{https://www.cosmos.esa.int/web/gaia/dpac/consortium}). Funding for the DPAC has been provided by national institutions, in particular the institutions participating in the {\it Gaia} Multilateral Agreement. This research has benefitted from the SpeX Prism Library (and/or SpeX Prism Library Analysis Toolkit), maintained by Adam Burgasser at \url{http://www.browndwarfs.org/spexprism}.  This research made use of the SIMBAD database and the VizieR Catalogue access tool, both operated at the CDS, Strasbourg, France. The original descriptions of the SIMBAD and VizieR services were published in \cite{2000A&AS..143....9W} and \cite{2000A&AS..143...23O}. This research has made use of NASA’s Astrophysics Data System Bibliographic Services.
\end{acknowledgements}

\bibliographystyle{aa}
\bibliography{bib.bib}

\begin{thebibliography}{62}
\expandafter\ifx\csname natexlab\endcsname\relax\def\natexlab#1{#1}\fi

\bibitem[{{Allard} {et~al.}(2012){Allard}, {Homeier}, \&
  {Freytag}}]{2012RSPTA.370.2765A}
{Allard}, F., {Homeier}, D., \& {Freytag}, B. 2012, Philosophical Transactions
  of the Royal Society of London Series A, 370, 2765

\bibitem[{{Amara} \& {Quanz}(2012)}]{2012MNRAS.427..948A}
{Amara}, A. \& {Quanz}, S.~P. 2012, \mnras, 427, 948

\bibitem[{Andrieu \& Thoms(2008)}]{andrieu2008}
Andrieu, C. \& Thoms, J. 2008, Statistics and Computing, 18, 343

\bibitem[{{Baraffe} {et~al.}(2003){Baraffe}, {Chabrier}, {Barman}, {Allard}, \&
  {Hauschildt}}]{2003A&A...402..701B}
{Baraffe}, I., {Chabrier}, G., {Barman}, T.~S., {Allard}, F., \& {Hauschildt},
  P.~H. 2003, \aap, 402, 701

\bibitem[{{Baraffe} {et~al.}(2015){Baraffe}, {Homeier}, {Allard}, \&
  {Chabrier}}]{2015A&A...577A..42B}
{Baraffe}, I., {Homeier}, D., {Allard}, F., \& {Chabrier}, G. 2015, \aap, 577,
  A42

\bibitem[{{Beuzit} {et~al.}(2019){Beuzit}, {Vigan}, {Mouillet}, {Dohlen},
  {Gratton}, {Boccaletti}, {Sauvage}, {Schmid}, {Langlois}, \&
  {Petit}}]{2019arXiv190204080B}
{Beuzit}, J.~L., {Vigan}, A., {Mouillet}, D., {et~al.} 2019, arXiv e-prints,
  arXiv:1902.04080

\bibitem[{{Boccaletti}(2019)}]{2019vltt.confE..38B}
{Boccaletti}, A. 2019, in The Very Large Telescope in 2030, 38

\bibitem[{{Boden} {et~al.}(2006){Boden}, {Torres}, \&
  {Latham}}]{2006ApJ...644.1193B}
{Boden}, A.~F., {Torres}, G., \& {Latham}, D.~W. 2006, \apj, 644, 1193

\bibitem[{{Bonnefoy} {et~al.}(2011){Bonnefoy}, {Lagrange}, {Boccaletti},
  {Chauvin}, {Apai}, {Allard}, {Ehrenreich}, {Girard}, {Mouillet}, {Rouan},
  {Gratadour}, \& {Kasper}}]{2011A&A...528L..15B}
{Bonnefoy}, M., {Lagrange}, A.~M., {Boccaletti}, A., {et~al.} 2011, \aap, 528,
  L15

\bibitem[{{Bowler} {et~al.}(2018){Bowler}, {Dupuy}, {Endl}, {Cochran},
  {MacQueen}, {Fulton}, {Petigura}, {Howard}, {Hirsch}, {Kratter}, {Crepp},
  {Biller}, {Johnson}, \& {Wittenmyer}}]{2018AJ....155..159B}
{Bowler}, B.~P., {Dupuy}, T.~J., {Endl}, M., {et~al.} 2018, \aj, 155, 159

\bibitem[{{Brandt} {et~al.}(2019){Brandt}, {Dupuy}, {Bowler}, {Bardalez
  Gagliuffi}, {Faherty}, {Mirek Brandt}, \& {Michalik}}]{2019arXiv191001652B}
{Brandt}, T.~D., {Dupuy}, T.~J., {Bowler}, B.~P., {et~al.} 2019, arXiv
  e-prints, arXiv:1910.01652

\bibitem[{{Burgasser}(2014)}]{2014ASInC..11....7B}
{Burgasser}, A.~J. 2014, in Astronomical Society of India Conference Series,
  Vol.~11, Astronomical Society of India Conference Series

\bibitem[{{Burgasser} {et~al.}(2016){Burgasser}, {Aganze}, {Escala}, {Lopez},
  {Choban}, {Jin}, {Iyer}, {Tallis}, {Suarez}, \& {Sahi}}]{2016AAS...22743408B}
{Burgasser}, A.~J., {Aganze}, C., {Escala}, I., {et~al.} 2016, in American
  Astronomical Society Meeting Abstracts, Vol. 227, American Astronomical
  Society Meeting Abstracts \#227, 434.08

\bibitem[{{Carlomagno} {et~al.}(2016){Carlomagno}, {Absil}, {Kenworthy},
  {Ruane}, {Keller}, {Otten}, {Feldt}, {Hippler}, {Huby}, {Mawet}, {Delacroix},
  {Surdej}, {Habraken}, {Forsberg}, {Karlsson}, {Vargas Catalan}, \&
  {Brandl}}]{2016SPIE.9909E..73C}
{Carlomagno}, B., {Absil}, O., {Kenworthy}, M., {et~al.} 2016, Society of
  Photo-Optical Instrumentation Engineers (SPIE) Conference Series, Vol. 9909,
  {End-to-end simulations of the E-ELT/METIS coronagraphs}, 990973

\bibitem[{{Cheetham} {et~al.}(2018){Cheetham}, {S{\'e}gransan}, {Peretti},
  {Delisle}, {Hagelberg}, {Beuzit}, {Forveille}, {Marmier}, {Udry}, \&
  {Wildi}}]{2018A&A...614A..16C}
{Cheetham}, A., {S{\'e}gransan}, D., {Peretti}, S., {et~al.} 2018, \aap, 614,
  A16

\bibitem[{{Crepp} {et~al.}(2016){Crepp}, {Gonzales}, {Bechter}, {Montet},
  {Johnson}, {Piskorz}, {Howard}, \& {Isaacson}}]{2016ApJ...831..136C}
{Crepp}, J.~R., {Gonzales}, E.~J., {Bechter}, E.~B., {et~al.} 2016, \apj, 831,
  136

\bibitem[{{Crepp} {et~al.}(2012){Crepp}, {Johnson}, {Fischer}, {Howard},
  {Marcy}, {Wright}, {Isaacson}, {Boyajian}, {von Braun}, {Hillenbrand},
  {Hinkley}, {Carpenter}, \& {Brewer}}]{2012ApJ...751...97C}
{Crepp}, J.~R., {Johnson}, J.~A., {Fischer}, D.~A., {et~al.} 2012, \apj, 751,
  97

\bibitem[{{Crepp} {et~al.}(2014){Crepp}, {Johnson}, {Howard}, {Marcy},
  {Brewer}, {Fischer}, {Wright}, \& {Isaacson}}]{2014ApJ...781...29C}
{Crepp}, J.~R., {Johnson}, J.~A., {Howard}, A.~W., {et~al.} 2014, \apj, 781, 29

\bibitem[{{Cutri} \& {et al.}(2013)}]{2013yCat.2328....0C}
{Cutri}, R.~M. \& {et al.} 2013, VizieR Online Data Catalog, II/328

\bibitem[{{Cutri} {et~al.}(2003){Cutri}, {Skrutskie}, {van Dyk}, {Beichman},
  {Carpenter}, {Chester}, {Cambresy}, {Evans}, {Fowler}, {Gizis}, {Howard},
  {Huchra}, {Jarrett}, {Kopan}, {Kirkpatrick}, {Light}, {Marsh}, {McCallon},
  {Schneider}, {Stiening}, {Sykes}, {Weinberg}, {Wheaton}, {Wheelock}, \&
  {Zacarias}}]{2003yCat.2246....0C}
{Cutri}, R.~M., {Skrutskie}, M.~F., {van Dyk}, S., {et~al.} 2003, VizieR Online
  Data Catalog, II/246

\bibitem[{{Delisle} {et~al.}(2016){Delisle}, {S{\'e}gransan}, {Buchschacher},
  \& {Alesina}}]{2016A&A...590A.134D}
{Delisle}, J.~B., {S{\'e}gransan}, D., {Buchschacher}, N., \& {Alesina}, F.
  2016, \aap, 590, A134

\bibitem[{{Delisle} {et~al.}(2018){Delisle}, {S{\'e}gransan}, {Dumusque},
  {Diaz}, {Bouchy}, {Lovis}, {Pepe}, {Udry}, {Alonso}, {Benz}, {Coffinet},
  {Collier Cameron}, {Deleuil}, {Figueira}, {Gillon}, {Lo Curto}, {Mayor},
  {Mordasini}, {Motalebi}, {Moutou}, {Pollacco}, {Pompei}, {Queloz}, {Santos},
  \& {Wyttenbach}}]{2018A&A...614A.133D}
{Delisle}, J.~B., {S{\'e}gransan}, D., {Dumusque}, X., {et~al.} 2018, \aap,
  614, A133

\bibitem[{{Dohlen} {et~al.}(2008){Dohlen}, {Langlois}, {Saisse}, {Hill},
  {Origne}, {Jacquet}, {Fabron}, {Blanc}, {Llored}, {Carle}, {Moutou}, {Vigan},
  {Boccaletti}, {Carbillet}, {Mouillet}, \& {Beuzit}}]{2008SPIE.7014E..3LD}
{Dohlen}, K., {Langlois}, M., {Saisse}, M., {et~al.} 2008, in \procspie, Vol.
  7014, Ground-based and Airborne Instrumentation for Astronomy II, 70143L

\bibitem[{{Dupuy} \& {Liu}(2017)}]{2017ApJS..231...15D}
{Dupuy}, T.~J. \& {Liu}, M.~C. 2017, \apjs, 231, 15

\bibitem[{{Ekstr{\"o}m} {et~al.}(2012){Ekstr{\"o}m}, {Georgy}, {Eggenberger},
  {Meynet}, {Mowlavi}, {Wyttenbach}, {Granada}, {Decressin}, {Hirschi},
  {Frischknecht}, {Charbonnel}, \& {Maeder}}]{2012A&A...537A.146E}
{Ekstr{\"o}m}, S., {Georgy}, C., {Eggenberger}, P., {et~al.} 2012, \aap, 537,
  A146

\bibitem[{{Feng} {et~al.}(2020){Feng}, {Butler}, {Shectman}, {Crane}, {Vogt},
  {Chambers}, {Jones}, {Xuesong Wang}, {Teske}, {Burt}, {Diaz}, \&
  {Thompson}}]{2020arXiv200102577F}
{Feng}, F., {Butler}, R.~P., {Shectman}, S.~A., {et~al.} 2020, arXiv e-prints,
  arXiv:2001.02577

\bibitem[{{Gaia Collaboration} {et~al.}(2018){Gaia Collaboration}, {Brown},
  {Vallenari}, {Prusti}, {de Bruijne}, {Babusiaux}, {Bailer-Jones}, {Biermann},
  {Evans}, {Eyer}, {Jansen}, {Jordi}, {Klioner}, {Lammers}, {Lindegren},
  {Luri}, {Mignard}, {Panem}, {Pourbaix}, {Randich}, {Sartoretti}, {Siddiqui},
  {Soubiran}, {van Leeuwen}, {Walton}, {Arenou}, {Bastian}, {Cropper},
  {Drimmel}, {Katz}, {Lattanzi}, {Bakker}, {Cacciari}, {Casta{\~n}eda},
  {Chaoul}, {Cheek}, {De Angeli}, {Fabricius}, {Guerra}, {Holl}, {Masana},
  {Messineo}, {Mowlavi}, {Nienartowicz}, {Panuzzo}, {Portell}, {Riello},
  {Seabroke}, {Tanga}, {Th{\'e}venin}, {Gracia-Abril}, {Comoretto},
  {Garcia-Reinaldos}, {Teyssier}, {Altmann}, {Andrae}, {Audard},
  {Bellas-Velidis}, {Benson}, {Berthier}, {Blomme}, {Burgess}, {Busso},
  {Carry}, {Cellino}, {Clementini}, {Clotet}, {Creevey}, {Davidson}, {De
  Ridder}, {Delchambre}, {Dell'Oro}, {Ducourant},
  {Fern{\'a}ndez-Hern{\'a}ndez}, {Fouesneau}, {Fr{\'e}mat}, {Galluccio},
  {Garc{\'\i}a-Torres}, {Gonz{\'a}lez-N{\'u}{\~n}ez}, {Gonz{\'a}lez-Vidal},
  {Gosset}, {Guy}, {Halbwachs}, {Hambly}, {Harrison}, {Hern{\'a}ndez},
  {Hestroffer}, {Hodgkin}, {Hutton}, {Jasniewicz}, {Jean-Antoine-Piccolo},
  {Jordan}, {Korn}, {Krone-Martins}, {Lanzafame}, {Lebzelter}, {L{\"o}ffler},
  {Manteiga}, {Marrese}, {Mart{\'\i}n-Fleitas}, {Moitinho}, {Mora}, {Muinonen},
  {Osinde}, {Pancino}, {Pauwels}, {Petit}, {Recio-Blanco}, {Richards},
  {Rimoldini}, {Robin}, {Sarro}, {Siopis}, {Smith}, {Sozzetti}, {S{\"u}veges},
  {Torra}, {van Reeven}, {Abbas}, {Abreu Aramburu}, {Accart}, {Aerts},
  {Altavilla}, {{\'A}lvarez}, {Alvarez}, {Alves}, {Anderson}, {Andrei},
  {Anglada Varela}, {Antiche}, {Antoja}, {Arcay}, {Astraatmadja}, {Bach},
  {Baker}, {Balaguer-N{\'u}{\~n}ez}, {Balm}, {Barache}, {Barata}, {Barbato},
  {Barblan}, {Barklem}, {Barrado}, {Barros}, {Barstow}, {Bartholom{\'e}
  Mu{\~n}oz}, {Bassilana}, {Becciani}, {Bellazzini}, {Berihuete}, {Bertone},
  {Bianchi}, {Bienaym{\'e}}, {Blanco-Cuaresma}, {Boch}, {Boeche}, {Bombrun},
  {Borrachero}, {Bossini}, {Bouquillon}, {Bourda}, {Bragaglia}, {Bramante},
  {Breddels}, {Bressan}, {Brouillet}, {Br{\"u}semeister}, {Brugaletta},
  {Bucciarelli}, {Burlacu}, {Busonero}, {Butkevich}, {Buzzi}, {Caffau},
  {Cancelliere}, {Cannizzaro}, {Cantat-Gaudin}, {Carballo}, {Carlucci},
  {Carrasco}, {Casamiquela}, {Castellani}, {Castro-Ginard}, {Charlot},
  {Chemin}, {Chiavassa}, {Cocozza}, {Costigan}, {Cowell}, {Crifo}, {Crosta},
  {Crowley}, {Cuypers}, {Dafonte}, {Damerdji}, {Dapergolas}, {David}, {David},
  {de Laverny}, {De Luise}, {De March}, {de Martino}, {de Souza}, {de Torres},
  {Debosscher}, {del Pozo}, {Delbo}, {Delgado}, {Delgado}, {Di Matteo},
  {Diakite}, {Diener}, {Distefano}, {Dolding}, {Drazinos}, {Dur{\'a}n},
  {Edvardsson}, {Enke}, {Eriksson}, {Esquej}, {Eynard Bontemps}, {Fabre},
  {Fabrizio}, {Faigler}, {Falc{\~a}o}, {Farr{\`a}s Casas}, {Federici},
  {Fedorets}, {Fernique}, {Figueras}, {Filippi}, {Findeisen}, {Fonti},
  {Fraile}, {Fraser}, {Fr{\'e}zouls}, {Gai}, {Galleti}, {Garabato},
  {Garc{\'\i}a-Sedano}, {Garofalo}, {Garralda}, {Gavel}, {Gavras}, {Gerssen},
  {Geyer}, {Giacobbe}, {Gilmore}, {Girona}, {Giuffrida}, {Glass}, {Gomes},
  {Granvik}, {Gueguen}, {Guerrier}, {Guiraud}, {Guti{\'e}rrez-S{\'a}nchez},
  {Haigron}, {Hatzidimitriou}, {Hauser}, {Haywood}, {Heiter}, {Helmi}, {Heu},
  {Hilger}, {Hobbs}, {Hofmann}, {Holland}, {Huckle}, {Hypki}, {Icardi},
  {Jan{\ss}en}, {Jevardat de Fombelle}, {Jonker}, {Juh{\'a}sz}, {Julbe},
  {Karampelas}, {Kewley}, {Klar}, {Kochoska}, {Kohley}, {Kolenberg},
  {Kontizas}, {Kontizas}, {Koposov}, {Kordopatis}, {Kostrzewa-Rutkowska},
  {Koubsky}, {Lambert}, {Lanza}, {Lasne}, {Lavigne}, {Le Fustec}, {Le
  Poncin-Lafitte}, {Lebreton}, {Leccia}, {Leclerc}, {Lecoeur-Taibi},
  {Lenhardt}, {Leroux}, {Liao}, {Licata}, {Lindstr{\o}m}, {Lister}, {Livanou},
  {Lobel}, {L{\'o}pez}, {Managau}, {Mann}, {Mantelet}, {Marchal}, {Marchant},
  {Marconi}, {Marinoni}, {Marschalk{\'o}}, {Marshall}, {Martino}, {Marton},
  {Mary}, {Massari}, {Matijevi{\v{c}}}, {Mazeh}, {McMillan}, {Messina},
  {Michalik}, {Millar}, {Molina}, {Molinaro}, {Moln{\'a}r}, {Montegriffo},
  {Mor}, {Morbidelli}, {Morel}, {Morris}, {Mulone}, {Muraveva}, {Musella},
  {Nelemans}, {Nicastro}, {Noval}, {O'Mullane}, {Ord{\'e}novic},
  {Ord{\'o}{\~n}ez-Blanco}, {Osborne}, {Pagani}, {Pagano}, {Pailler},
  {Palacin}, {Palaversa}, {Panahi}, {Pawlak}, {Piersimoni}, {Pineau}, {Plachy},
  {Plum}, {Poggio}, {Poujoulet}, {Pr{\v{s}}a}, {Pulone}, {Racero}, {Ragaini},
  {Rambaux}, {Ramos-Lerate}, {Regibo}, {Reyl{\'e}}, {Riclet}, {Ripepi}, {Riva},
  {Rivard}, {Rixon}, {Roegiers}, {Roelens}, {Romero-G{\'o}mez}, {Rowell},
  {Royer}, {Ruiz-Dern}, {Sadowski}, {Sagrist{\`a} Sell{\'e}s}, {Sahlmann},
  {Salgado}, {Salguero}, {Sanna}, {Santana-Ros}, {Sarasso}, {Savietto},
  {Schultheis}, {Sciacca}, {Segol}, {Segovia}, {S{\'e}gransan}, {Shih},
  {Siltala}, {Silva}, {Smart}, {Smith}, {Solano}, {Solitro}, {Sordo}, {Soria
  Nieto}, {Souchay}, {Spagna}, {Spoto}, {Stampa}, {Steele},
  {Steidelm{\"u}ller}, {Stephenson}, {Stoev}, {Suess}, {Surdej}, {Szabados},
  {Szegedi-Elek}, {Tapiador}, {Taris}, {Tauran}, {Taylor}, {Teixeira},
  {Terrett}, {Teyssand ier}, {Thuillot}, {Titarenko}, {Torra Clotet}, {Turon},
  {Ulla}, {Utrilla}, {Uzzi}, {Vaillant}, {Valentini}, {Valette}, {van Elteren},
  {Van Hemelryck}, {van Leeuwen}, {Vaschetto}, {Vecchiato}, {Veljanoski},
  {Viala}, {Vicente}, {Vogt}, {von Essen}, {Voss}, {Votruba}, {Voutsinas},
  {Walmsley}, {Weiler}, {Wertz}, {Wevers}, {Wyrzykowski}, {Yoldas},
  {{\v{Z}}erjal}, {Ziaeepour}, {Zorec}, {Zschocke}, {Zucker}, {Zurbach}, \&
  {Zwitter}}]{2018A&A...616A...1G}
{Gaia Collaboration}, {Brown}, A.~G.~A., {Vallenari}, A., {et~al.} 2018, \aap,
  616, A1

\bibitem[{Gao \& Han(2012)}]{Gao2012}
Gao, F. \& Han, L. 2012, Computational Optimization and Applications, 51, 259

\bibitem[{{Georgy} {et~al.}(2013){Georgy}, {Ekstr{\"o}m}, {Eggenberger},
  {Meynet}, {Haemmerl{\'e}}, {Maeder}, {Granada}, {Groh}, {Hirschi}, {Mowlavi},
  {Yusof}, {Charbonnel}, {Decressin}, \& {Barblan}}]{2013A&A...558A.103G}
{Georgy}, C., {Ekstr{\"o}m}, S., {Eggenberger}, P., {et~al.} 2013, \aap, 558,
  A103

\bibitem[{Haario {et~al.}(2001)Haario, Saksman, \& Tamminen}]{haario2001}
Haario, H., Saksman, E., \& Tamminen, J. 2001, Bernoulli, 7, 223

\bibitem[{{Hagelberg} {et~al.}(2016){Hagelberg}, {S{\'e}gransan}, {Udry}, \&
  {Wildi}}]{2016MNRAS.455.2178H}
{Hagelberg}, J., {S{\'e}gransan}, D., {Udry}, S., \& {Wildi}, F. 2016, \mnras,
  455, 2178

\bibitem[{{Hoeg} {et~al.}(1997){Hoeg}, {B{\"a}ssgen}, {Bastian}, {Egret},
  {Fabricius}, {Gro{\ss}mann}, {Halbwachs}, {Makarov}, {Perryman},
  {Schwekendiek}, {Wagner}, \& {Wicenec}}]{1997A&A...323L..57H}
{Hoeg}, E., {B{\"a}ssgen}, G., {Bastian}, U., {et~al.} 1997, \aap, 323, L57

\bibitem[{{H{\o}g} {et~al.}(2000){H{\o}g}, {Fabricius}, {Makarov}, {Urban},
  {Corbin}, {Wycoff}, {Bastian}, {Schwekendiek}, \&
  {Wicenec}}]{2000A&A...355L..27H}
{H{\o}g}, E., {Fabricius}, C., {Makarov}, V.~V., {et~al.} 2000, \aap, 355, L27

\bibitem[{{Kirkpatrick} {et~al.}(2000){Kirkpatrick}, {Reid}, {Liebert},
  {Gizis}, {Burgasser}, {Monet}, {Dahn}, {Nelson}, \&
  {Williams}}]{2000AJ....120..447K}
{Kirkpatrick}, J.~D., {Reid}, I.~N., {Liebert}, J., {et~al.} 2000, \aj, 120,
  447

\bibitem[{{Liu} {et~al.}(2002){Liu}, {Fischer}, {Graham}, {Lloyd}, {Marcy}, \&
  {Butler}}]{2002ApJ...571..519L}
{Liu}, M.~C., {Fischer}, D.~A., {Graham}, J.~R., {et~al.} 2002, \apj, 571, 519

\bibitem[{{Maire} {et~al.}(2019){Maire}, {Baudino}, {Desidera}, {Messina},
  {Brandner}, {Godoy}, {Cantalloube}, {Galicher}, {Bonnefoy}, {Hagelberg},
  {Olofsson}, {Absil}, {Chauvin}, {Henning}, \&
  {Langlois}}]{2019arXiv191202565M}
{Maire}, A.~L., {Baudino}, J.~L., {Desidera}, S., {et~al.} 2019, arXiv
  e-prints, arXiv:1912.02565

\bibitem[{{Maire} {et~al.}(2016{\natexlab{a}}){Maire}, {Bonnefoy}, {Ginski},
  {Vigan}, {Messina}, {Mesa}, {Galicher}, {Gratton}, {Desidera}, {Kopytova},
  {Millward}, {Thalmann}, {Claudi}, {Ehrenreich}, {Zurlo}, {Chauvin},
  {Antichi}, {Baruffolo}, {Bazzon}, {Beuzit}, {Blanchard}, {Boccaletti}, {de
  Boer}, {Carle}, {Cascone}, {Costille}, {De Caprio}, {Delboulb{\'e}},
  {Dohlen}, {Dominik}, {Feldt}, {Fusco}, {Girard}, {Giro}, {Gisler}, {Gluck},
  {Gry}, {Henning}, {Hubin}, {Hugot}, {Jaquet}, {Kasper}, {Lagrange},
  {Langlois}, {Le Mignant}, {Llored}, {Madec}, {Martinez}, {Mawet}, {Milli},
  {M{\"o}ller-Nilsson}, {Mouillet}, {Moulin}, {Moutou}, {Orign{\'e}}, {Pavlov},
  {Petit}, {Pragt}, {Puget}, {Ramos}, {Rochat}, {Roelfsema}, {Salasnich},
  {Sauvage}, {Schmid}, {Turatto}, {Udry}, {Vakili}, {Wahhaj}, {Weber}, \&
  {Wildi}}]{2016A&A...587A..56M}
{Maire}, A.~L., {Bonnefoy}, M., {Ginski}, C., {et~al.} 2016{\natexlab{a}},
  \aap, 587, A56

\bibitem[{{Maire} {et~al.}(2016{\natexlab{b}}){Maire}, {Langlois}, {Dohlen},
  {Lagrange}, {Gratton}, {Chauvin}, {Desidera}, {Girard}, {Milli}, {Vigan},
  {Zins}, {Delorme}, {Beuzit}, {Claudi}, {Feldt}, {Mouillet}, {Puget},
  {Turatto}, \& {Wildi}}]{2016SPIE.9908E..34M}
{Maire}, A.-L., {Langlois}, M., {Dohlen}, K., {et~al.} 2016{\natexlab{b}},
  Society of Photo-Optical Instrumentation Engineers (SPIE) Conference Series,
  Vol. 9908, {SPHERE IRDIS and IFS astrometric strategy and calibration},
  990834

\bibitem[{{Mamajek} \& {Hillenbrand}(2008)}]{2008ApJ...687.1264M}
{Mamajek}, E.~E. \& {Hillenbrand}, L.~A. 2008, \apj, 687, 1264

\bibitem[{{Marley} {et~al.}(1996){Marley}, {Saumon}, {Guillot}, {Freedman},
  {Hubbard}, {Burrows}, \& {Lunine}}]{1996Sci...272.1919M}
{Marley}, M.~S., {Saumon}, D., {Guillot}, T., {et~al.} 1996, Science, 272, 1919

\bibitem[{{Marmier}(2014)}]{marmier_phd_thesis}
{Marmier}, M. 2014, PhD thesis, Geneva Observatory, University of Geneva,
  Switzerland

\bibitem[{{Mayor} {et~al.}(2003){Mayor}, {Pepe}, {Queloz}, {Bouchy},
  {Rupprecht}, {Lo Curto}, {Avila}, {Benz}, {Bertaux}, {Bonfils}, {Dall},
  {Dekker}, {Delabre}, {Eckert}, {Fleury}, {Gilliotte}, {Gojak}, {Guzman},
  {Kohler}, {Lizon}, {Longinotti}, {Lovis}, {Megevand}, {Pasquini}, {Reyes},
  {Sivan}, {Sosnowska}, {Soto}, {Udry}, {van Kesteren}, {Weber}, \&
  {Weilenmann}}]{2003Msngr.114...20M}
{Mayor}, M., {Pepe}, F., {Queloz}, D., {et~al.} 2003, The Messenger, 114, 20

\bibitem[{{Mesa} {et~al.}(2015){Mesa}, {Gratton}, {Zurlo}, {Vigan}, {Claudi},
  {Alberi}, {Antichi}, {Baruffolo}, {Beuzit}, {Boccaletti}, {Bonnefoy},
  {Costille}, {Desidera}, {Dohlen}, {Fantinel}, {Feldt}, {Fusco}, {Giro},
  {Henning}, {Kasper}, {Langlois}, {Maire}, {Martinez}, {Moeller-Nilsson},
  {Mouillet}, {Moutou}, {Pavlov}, {Puget}, {Salasnich}, {Sauvage}, {Sissa},
  {Turatto}, {Udry}, {Vakili}, {Waters}, \& {Wildi}}]{2015A&A...576A.121M}
{Mesa}, D., {Gratton}, R., {Zurlo}, A., {et~al.} 2015, \aap, 576, A121

\bibitem[{{Morley} {et~al.}(2012){Morley}, {Fortney}, {Marley}, {Visscher},
  {Saumon}, \& {Leggett}}]{2012ApJ...756..172M}
{Morley}, C.~V., {Fortney}, J.~J., {Marley}, M.~S., {et~al.} 2012, \apj, 756,
  172

\bibitem[{{Nakajima} {et~al.}(1995){Nakajima}, {Oppenheimer}, {Kulkarni},
  {Golimowski}, {Matthews}, \& {Durrance}}]{1995Natur.378..463N}
{Nakajima}, T., {Oppenheimer}, B.~R., {Kulkarni}, S.~R., {et~al.} 1995, \nat,
  378, 463

\bibitem[{{Ochsenbein} {et~al.}(2000){Ochsenbein}, {Bauer}, \&
  {Marcout}}]{2000A&AS..143...23O}
{Ochsenbein}, F., {Bauer}, P., \& {Marcout}, J. 2000, \aaps, 143, 23

\bibitem[{{Pavlov} {et~al.}(2008){Pavlov}, {M{\"o}ller-Nilsson}, {Feldt},
  {Henning}, {Beuzit}, \& {Mouillet}}]{2008SPIE.7019E..39P}
{Pavlov}, A., {M{\"o}ller-Nilsson}, O., {Feldt}, M., {et~al.} 2008, Society of
  Photo-Optical Instrumentation Engineers (SPIE) Conference Series, Vol. 7019,
  {SPHERE data reduction and handling system: overview, project status, and
  development}, 701939

\bibitem[{{Peretti} {et~al.}(2019){Peretti}, {S{\'e}gransan}, {Lavie},
  {Desidera}, {Maire}, {D'Orazi}, {Vigan}, {Baudino}, {Cheetham}, {Janson},
  {Chauvin}, {Hagelberg}, {Menard}, {Heng}, {Udry}, {Boccaletti}, {Daemgen},
  {Le Coroller}, {Mesa}, {Rouan}, {Samland}, {Schmidt}, {Zurlo}, {Bonnefoy},
  {Feldt}, {Gratton}, {Lagrange}, {Langlois}, {Meyer}, {Carbillet}, {Carle},
  {De Caprio}, {Gluck}, {Hugot}, {Magnard}, {Moulin}, {Pavlov}, {Pragt},
  {Rabou}, {Ramos}, {Rousset}, {Sevin}, {Soenke}, {Stadler}, {Weber}, \&
  {Wildi}}]{2019A&A...631A.107P}
{Peretti}, S., {S{\'e}gransan}, D., {Lavie}, B., {et~al.} 2019, \aap, 631, A107

\bibitem[{{Perryman} {et~al.}(1997){Perryman}, {Lindegren}, {Kovalevsky},
  {Hoeg}, {Bastian}, {Bernacca}, {Cr{\'e}z{\'e}}, {Donati}, {Grenon},
  {Grewing}, {van Leeuwen}, {van der Marel}, {Mignard}, {Murray}, {Le Poole},
  {Schrijver}, {Turon}, {Arenou}, {Froeschl{\'e}}, \&
  {Petersen}}]{1997A&A...323L..49P}
{Perryman}, M.~A.~C., {Lindegren}, L., {Kovalevsky}, J., {et~al.} 1997, \aap,
  323, L49

\bibitem[{{Queloz} {et~al.}(2000){Queloz}, {Mayor}, {Naef}, {Santos}, {Udry},
  {Burnet}, \& {Confino}}]{2000fepc.conf..548Q}
{Queloz}, D., {Mayor}, M., {Naef}, D., {et~al.} 2000, in From Extrasolar
  Planets to Cosmology: The VLT Opening Symposium, ed. J.~{Bergeron} \&
  A.~{Renzini}, 548

\bibitem[{{Rickman} {et~al.}(2019){Rickman}, {S{\'e}gransan}, {Marmier},
  {Udry}, {Bouchy}, {Lovis}, {Mayor}, {Pepe}, {Queloz}, \&
  {Santos}}]{2019A&A...625A..71R}
{Rickman}, E.~L., {S{\'e}gransan}, D., {Marmier}, M., {et~al.} 2019, \aap, 625,
  A71

\bibitem[{{Sahlmann} {et~al.}(2011){Sahlmann}, {S{\'e}gransan}, {Queloz},
  {Udry}, {Santos}, {Marmier}, {Mayor}, {Naef}, {Pepe}, \&
  {Zucker}}]{2011A&A...525A..95S}
{Sahlmann}, J., {S{\'e}gransan}, D., {Queloz}, D., {et~al.} 2011, \aap, 525,
  A95

\bibitem[{{Santos} {et~al.}(2001){Santos}, {Israelian}, \&
  {Mayor}}]{2001A&A...373.1019S}
{Santos}, N.~C., {Israelian}, G., \& {Mayor}, M. 2001, \aap, 373, 1019

\bibitem[{{Santos} {et~al.}(2013){Santos}, {Sousa}, {Mortier}, {Neves},
  {Adibekyan}, {Tsantaki}, {Delgado Mena}, {Bonfils}, {Israelian}, {Mayor}, \&
  {Udry}}]{2013A&A...556A.150S}
{Santos}, N.~C., {Sousa}, S.~G., {Mortier}, A., {et~al.} 2013, \aap, 556, A150

\bibitem[{{S{\'e}gransan} {et~al.}(2010){S{\'e}gransan}, {Udry}, {Mayor},
  {Naef}, {Pepe}, {Queloz}, {Santos}, {Demory}, {Figueira}, {Gillon},
  {Marmier}, {M{\'e}gevand}, {Sosnowska}, {Tamuz}, \&
  {Triaud}}]{2010A&A...511A..45S}
{S{\'e}gransan}, D., {Udry}, S., {Mayor}, M., {et~al.} 2010, \aap, 511, A45

\bibitem[{{Soummer} {et~al.}(2003){Soummer}, {Aime}, \&
  {Falloon}}]{2003A&A...397.1161S}
{Soummer}, R., {Aime}, C., \& {Falloon}, P.~E. 2003, \aap, 397, 1161

\bibitem[{{Soummer} {et~al.}(2012){Soummer}, {Pueyo}, \&
  {Larkin}}]{2012ApJ...755L..28S}
{Soummer}, R., {Pueyo}, L., \& {Larkin}, J. 2012, \apjl, 755, L28

\bibitem[{{Thalmann} {et~al.}(2009){Thalmann}, {Carson}, {Janson}, {Goto},
  {McElwain}, {Egner}, {Feldt}, {Hashimoto}, {Hayano}, {Henning}, {Hodapp},
  {Kandori}, {Klahr}, {Kudo}, {Kusakabe}, {Mordasini}, {Morino}, {Suto},
  {Suzuki}, \& {Tamura}}]{2009ApJ...707L.123T}
{Thalmann}, C., {Carson}, J., {Janson}, M., {et~al.} 2009, \apjl, 707, L123

\bibitem[{{Udry} {et~al.}(2000){Udry}, {Mayor}, {Queloz}, {Naef}, \&
  {Santos}}]{2000fepc.conf..571U}
{Udry}, S., {Mayor}, M., {Queloz}, D., {Naef}, D., \& {Santos}, N. 2000, in
  From Extrasolar Planets to Cosmology: The VLT Opening Symposium, ed.
  J.~{Bergeron} \& A.~{Renzini}, 571

\bibitem[{{Vigan} {et~al.}(2016){Vigan}, {Bonnefoy}, {Ginski}, {Beust},
  {Galicher}, {Janson}, {Baudino}, {Buenzli}, {Hagelberg}, {D'Orazi},
  {Desidera}, {Maire}, {Gratton}, {Sauvage}, {Chauvin}, {Thalmann}, {Malo},
  {Salter}, {Zurlo}, {Antichi}, {Baruffolo}, {Baudoz}, {Blanchard},
  {Boccaletti}, {Beuzit}, {Carle}, {Claudi}, {Costille}, {Delboulb{\'e}},
  {Dohlen}, {Dominik}, {Feldt}, {Fusco}, {Gluck}, {Girard}, {Giro}, {Gry},
  {Henning}, {Hubin}, {Hugot}, {Jaquet}, {Kasper}, {Lagrange}, {Langlois}, {Le
  Mignant}, {Llored}, {Madec}, {Martinez}, {Mawet}, {Mesa}, {Milli},
  {Mouillet}, {Moulin}, {Moutou}, {Orign{\'e}}, {Pavlov}, {Perret}, {Petit},
  {Pragt}, {Puget}, {Rabou}, {Rochat}, {Roelfsema}, {Salasnich}, {Schmid},
  {Sevin}, {Siebenmorgen}, {Smette}, {Stadler}, {Suarez}, {Turatto}, {Udry},
  {Vakili}, {Wahhaj}, {Weber}, \& {Wildi}}]{2016A&A...587A..55V}
{Vigan}, A., {Bonnefoy}, M., {Ginski}, C., {et~al.} 2016, \aap, 587, A55

\bibitem[{{Vigan} {et~al.}(2010){Vigan}, {Moutou}, {Langlois}, {Allard},
  {Boccaletti}, {Carbillet}, {Mouillet}, \& {Smith}}]{2010MNRAS.407...71V}
{Vigan}, A., {Moutou}, C., {Langlois}, M., {et~al.} 2010, \mnras, 407, 71

\bibitem[{{Wenger} {et~al.}(2000){Wenger}, {Ochsenbein}, {Egret}, {Dubois},
  {Bonnarel}, {Borde}, {Genova}, {Jasniewicz}, {Lalo{\"e}}, {Lesteven}, \&
  {Monier}}]{2000A&AS..143....9W}
{Wenger}, M., {Ochsenbein}, F., {Egret}, D., {et~al.} 2000, \aaps, 143, 9

\end{thebibliography}

\appendix
\section{MCMC results}

Here we provide the full list of parameters adjusted in the MCMC as well as the marginalised 1D and 2D posterior distributions of the parameters corresponding to the global fit of the RV and direct-imaging models.

\begin{sidewaystable*}
    \caption{Parameters corresponding to the combined adjustment of the astrometric and RV observations of HD13724~AB derived from the MCMC posterior distributions.}
    \centering
    \begin{tabular}{ccccccccc}
    \hline
    \hline
     Var                 &  Units  &    Max(Proba)   &      Mode    &       Med   &    Std  &  CI(2.5\%)  &    CI(97.5\%) & Priors \\       

M$_{\star}$                    &  [M$_\odot$] &       &        &     &    &    &      &$N(1.14, 0.06)$ \\         
$\pi$                    & [mas]  &       &        &     &    &    &      &$N(22.9781,\sqrt{0.0277^2 + 0.040^2})$ \\   
\hline                                                                                         
$\log{(\rm Proba)}$            &         &   -675.516   &   -685.396   &   -682.013  &   2.687 &  -688.373  &    -678.024 & \\       
\hline                                                                                         
$\gamma(H03$)            &  [m/s]  &  20833   &  20885   &  20806  &  36 & 20733  &   20870 &$U$\\       
$\Delta V(CVL-H03)^{*}$  &  [m/s]  &   -229   &   -303   &   -234  & 133 &  -496  &      27 &$N(0,300)$\\       
$\Delta V(C98-H03)$      &  [m/s]  &    -37.9   &    -35.8   &    -37.6  &   2.3 &   -42.2  &     -33.0 &$U$\\       
$\Delta V(C07-H03)$      &  [m/s]  &    -37.8   &    -34.2   &    -41.0  &   4.5 &   -49.9  &     -32.2 &$U$\\       
$\Delta V(C14-H03)$      &  [m/s]  &    -15.1   &    -21.2   &    -13.6  &   4.4 &   -22.3  &      -5.0 &$N(-20, 5)$\\       
$S_{\rm Act}$            &  [m/s]  &     14.0   &     20.3   &     12.5  &   4.5 &     3.6  &      21.4 &$U$\\    
$\sigma_J$               &  [m/s]  &     7.03   &      7.17   &      7.11  &   0.62 &     5.98  &       8.39 &$U$\\       
\hline 
$\log{(P)}$              &  [day]  &      4.764   &      4.952   &      4.654  &   0.132 &     4.423  &       4.917 &$U$\\       
$\log{(K)}$              &  [m/s]  &      2.459   &      2.483   &      2.444  &   0.020 &     2.400  &       2.476 &$U$\\       
$\sqrt{e}.\cos{\omega}$  &         &     -0.827   &     -0.864   &     -0.792  &   0.043 &    -0.862  &      -0.702 &$U$\\      
$\sqrt{e}.\sin{\omega}$  &         &     -0.070   &     -0.114   &     -0.067  &   0.014 &    -0.095  &      -0.039 &$U$\\      
$T_{\rm Vmin}$           &  [bjd]$^{**}$  &  56033   &  56038   &  56034  &  21 & 55993  &   56074&$U$ \\       
$\Omega$                 &  [deg]  &     17.4   &     19.2   &     16.4  &   4.3 &     7.3  &      23.9 &$U$ \\        
$i$                      &  [deg]  &     58.5   &     58.6   &     57.40  &   1.7 &    53.6  &      60.0 &$U$ \\         
$a_r$                    &  [mas]  &    721   &    945   &    606  & 129 &   421  &     907 &$U$ \\         

\hline                                                                                         
K                        &  [m/s]  &    288  &     304   &    278  &  13 &   251&     299& - \\
P                        &    [y]  &    159  &     245  &    123  &  41 &    72 &     226&  -\\
$e$                      &         &      0.688  &       0.760   &      0.632  &   0.067 &     0.498 &       0.747& - \\
 $\omega$                &  [deg]  &   184.9& 187.5 & 184.9& 1.1& 182.8& 187.0& -\\
$a_r$                    &   [au]  &     31.3  &      41.2   &     26.3  &   5.6 &    18.3 &      39.5 & -\\
M$_B$                    & [Mjup]  &     53.1  &      56.3   &     50.5  &   3.3 &    43.7 &      56.4& - \\
M$_B.\sin(i)$            & [Mjup]  &     45.3  &      48.1   &     42.5  &   3.5 &    35.3 &      48.6 & -\\
 \hline 
         \multicolumn{8}{l}{*: CVL stands for CORAVEL; **: The date is expressed as BJD-2400000.}\\
    \end{tabular}
    \label{tab:ds2}
\end{sidewaystable*}

\begin{figure*}
\includegraphics[width=0.99\textwidth]{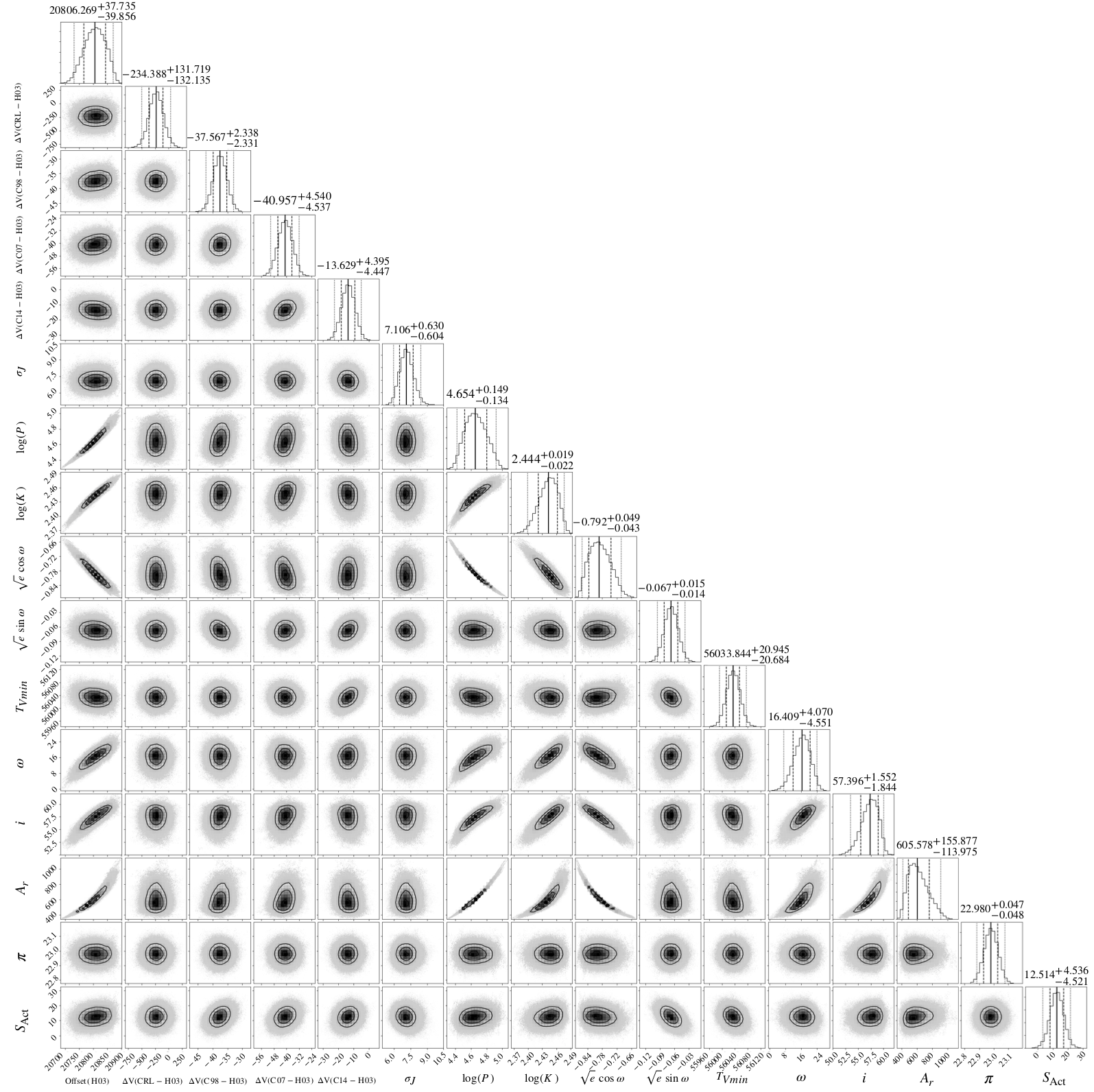}
\caption{Marginalised 1D and 2D posterior distributions of the 16 parameters corresponding to the global fit of the RV and direct-imaging models. Confidence intervals at 2.275\%, 15.85\%, 50.0\%, 84.15\%, and 97.725\% are overplotted on the 1D posterior distributions, while the median $\pm 1~\sigma$ values are given at the top of each 1D distribution. Overplotted on the 2D posterior distribution are the 1, 2, and 3~$\sigma$ contour levels. 
From left-to-right and top-to-bottom, the list of parameters are: 
$\gamma(H03)$, $\Delta V(CVL-H03)$, $\Delta V(C98-H03)$, $\Delta V(C07-H03)$, $\Delta V(C14-H03)$, $\sigma_J$, $\log{(P)}$, $\log{(K)}$, $\sqrt{e}.\cos{\omega}$, $\sqrt{e}.\sin{\omega}$, $T_{\rm Vmin}$, $\Omega$, $i$, $a_r$, $\pi$, $S_{\rm Act}$.}
\label{fig:A1}
\end{figure*}

\end{document}